\documentclass[preprint,superscriptaddress]{article}
\usepackage[utf8]{inputenc}
\usepackage{amssymb}
\usepackage{amsmath,esint}
\usepackage{gensymb}
\usepackage{svg}
\usepackage{graphicx}
\usepackage{siunitx}
\usepackage{hyperref}
\usepackage[version=4]{mhchem}
\usepackage[margin=1in]{geometry}
\usepackage{xcolor}

\usepackage[square, numbers, sort&compress]{natbib}
\newcommand{\Rth}{$R_{th}$}
\newcommand{\CrBST}{$\mathrm{Cr_{0.15}(Bi,Sb)_{1.85}Te_{3}}$}
\newcommand{\BST}{$\mathrm{(Bi,Sb)_{1.85}Te_{3}}$}

\newcommand{\STO}{SrTiO$\mathrm{_3}$}

\begin{document}
\title{Visualizing the breakdown of the quantum anomalous Hall effect}

\author{
G. M. Ferguson$^{1}$, Run Xiao$^{2}$, Anthony R. Richardella$^{2}$, Austin Kaczmarek$^{1}$\\ Nitin Samarth$^{2}$, Katja C. Nowack$^{1,3}$*
}

\date{
\normalsize{$^{1}$Laboratory of Atomic and Solid-State Physics, Cornell University, Ithaca, NY 14853, USA} 
\\
\normalsize{$^{2}$Department of Physics, The Pennsylvania State University, University Park, 16802, Pennsylvania, USA}
\\
\normalsize{$^{3}$Kavli Institute at Cornell for Nanoscale Science, Cornell University, Ithaca, NY 14853, USA}
\\
\normalsize{$^\ast$To whom correspondence should be addressed; email: kcn34@cornell.edu}}

\maketitle
\begin{abstract}
\textbf{
The creation of topologically non-trivial matter across electronic, mechanical, cold-atom, and photonic platforms is advancing rapidly, yet understanding the breakdown of topological protection remains a major challenge. In this work, we use magnetic imaging combined with global electrical transport measurements to visualize the current-induced breakdown of the quantum anomalous Hall effect (QAHE) in a magnetically doped topological insulator. We find that dissipation emerges at localized hot spots near electrical contacts, where an abrupt change in Hall angle leads to significant distortions of the current density. Using the local magnetization as a proxy for electron temperature, we directly observe that the electrons are driven out of equilibrium with the lattice at the hot spots and throughout the device in the breakdown regime. By characterizing energy relaxation processes in our device, we show that the breakdown of quantization is governed entirely by electron heating, and that a vanishing thermal relaxation strength at millikelvin temperatures limits the robustness of the QAHE. Our findings provide a framework for diagnosing energy relaxation in topological materials and will guide realizing robust topological protection in magnetic topological insulators.  
}
\end{abstract}

\newpage

\maketitle
The quantum anomalous Hall effect (QAHE), characterized by a quantized Hall effect and a vanishing longitudinal resistance at zero magnetic field, is an example of a topologically protected state of matter \cite{chang2023review}. Although the QAHE has potential applications in metrology \cite{okazaki2022quantum, rodenbach2022metrological} as well as classical \cite{fan2014magnetization, fan2016electric, kondou2016fermi, yuan2024electrical} and quantum information processing \cite{lian2018topological}, experimental realizations of the QAHE remain limited to cryogenic temperatures and low bias currents. Understanding the microscopic origin of these limitations to the QAHE is therefore crucial for realizing a robust QAHE at higher temperatures and bias currents. 

\begin{figure}
    \centering
    \includegraphics[width=1.0\textwidth]{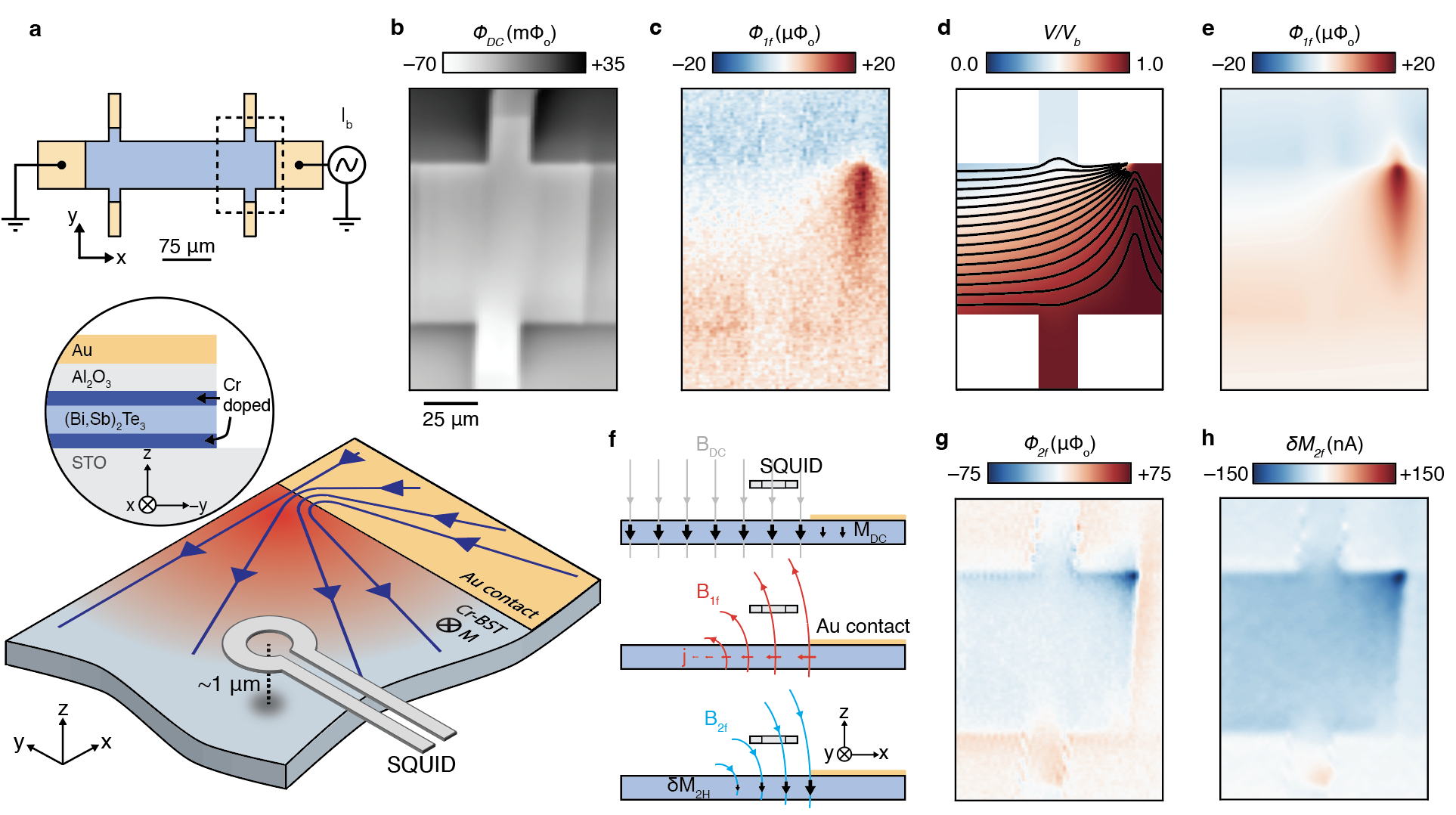}
    \caption{\textbf{Imaging currents and heating in a quantum anomalous Hall insulator. a,} Schematic of the measurement showing (top) the sample geometry and biasing scheme with the dashed box indicating the field of view, (middle) a cross-section of the magnetically doped topological insulator heterostructure, and (bottom) an illustration of the hot-spot region  where current enters at the contact corner causing high local power dissipation. A SQUID with a micrometer-scale pickup loop detects magnetic signals from the sample.  \textbf{b,} Image of the DC flux, $\Phi_{DC}$, measured  in the field of view indicated in (\textbf{a}, top) with the sample magnetized into the plane. \textbf{c,} First harmonic SQUID signal, $\Phi_{1f}$, co-recorded with (\textbf{b}). The source-drain bias current $I_b$ is \SI{120}{\nano\ampere}. \textbf{d,} Electrostatic potential distribution calculated using the measured longitudinal and Hall resistivities. Black lines show current streamlines calculated from the potential and the resistivity tensor. \textbf{e,} Model magnetic flux image obtained by convolving the the model current distribution in (\textbf{d}) with the SQUID's imaging kernel. \textbf{f,} Schematic illustrating the three modes of magnetic imaging demonstrated in (\textbf{b}), (\textbf{c}) and (\textbf{g}). The stray field from the sample magnetization couples a DC flux $\Phi_{DC}$ into the SQUID pickup loop (top). The transport current distribution couples a flux $\Phi_{1f}$ into the SQUID at the at the frequency of the excitation (middle). Changes in the sample magnetization due to heating in the sample couples a flux $\Phi_{2f}$ into the SQUID at twice the frequency of the excitation (bottom). \textbf{g,} Second harmonic SQUID signal, $\Phi_{2f}$ co-recorded with (\textbf{c}). \textbf{h,} Change in the sample magnetization $\delta M_{2f}$ induced by bias-induced heating, reconstructed from (\textbf{g}). Data shown here is collected at base temperature of our dilution refrigerator, $\sim\SI{15}{\milli\kelvin}$.}
\end{figure}

At sufficiently high temperatures or bias currents, all QAH systems undergo a breakdown process marked by the deviation of the electronic transport coefficients from their quantized values. Electrical transport measurements of breakdown have focused on magnetically doped topological insulators (TIs) in which the QAHE has been first realized \cite{chang2013experimental}. The results have been interpreted using a range of models, including field-assisted variable range hopping \cite{kawamura2017current}, bootstrap electron heating \cite{fox2018part} and electric field driven breakdown \cite{lippertz2022current}. Although QAH systems are predicted to host topologically protected edge states, the role that these edge states play in the transport and breakdown of the QAHE remains controversial \cite{chang2015zero, lippertz2022current, rodenbach2021bulk,rosen2022measured,ferguson2023direct,doucot2024meandering}. Despite this interest in understanding the breakdown of the QAHE, its microscopic nature remains unclear.

In this work, we use magnetic imaging combined with global electrical transport measurements to characterize the current-induced breakdown of the quantum anomalous Hall regime in the magnetically doped TI \CrBST{}. We find that the magnetization serves as a sensor of the local electron temperature, providing micrometer-scale information about where power is dissipated in our device. Simultaneously, we acquire images of the stray magnetic field produced by the current in the device, allowing us to determine the current distribution. For magnetic imaging, we use a scanning superconducting quantum interference device (SQUID) microscope \cite{low2021scanning}, illustrated schematically in Fig. 1a. We approach a SQUID with a micrometer-scale pickup loop to the surface of a \CrBST{} Hall bar. The device is fabricated on a \ce{SrTiO3} substrate, which serves as a global back gate, and includes a top gate that was grounded for the measurements described here. Characterization of the electronic transport behavior and the current distribution in the channel in the low-bias limit was presented in Ref. \cite{ferguson2023direct} as Device C. To compare our sample with previous work on the breakdown of the QAHE, we present the temperature-dependent conductivity of our device in Supplementary Information Section 1.

In our measurements, several sources of stray magnetic fields are present. The static magnetization of the Cr-dopants produces a static field, which we detect as a DC signal, $\Phi_{DC}$. We apply a bias current $I_b$ at a frequency $f = \SI{140.6}{\hertz}$ through the channel of the sample. The resulting alternating current generates a stray magnetic field at the same frequency. We detect the corresponding flux, $\Phi_{1f}$, coupled into the SQUID pickup loop using lock-in amplification. In addition, we observe a second harmonic response,  $\Phi_{2f}$, at twice the excitation frequency, which we show below arises from small changes in the sample magnetization due to current-induced heating. In Fig. 1b, we present a $\Phi_{DC}$ image acquired after magnetizing the sample into the plane ($-z$). The $\Phi_{DC}$ image reveals the outline of the channel and two of the voltage probes. Figs. 1c and 1g show the simultaneously acquired $\Phi_{1f}$ and $\Phi_{2f}$ signals. The origins of $\Phi_{DC}$, $\Phi_{1f}$ and $\Phi_{2f}$ signals are illustrated schematically in Fig. 1h. Further details on the sample fabrication, scanning SQUID measurements and lock-in detection scheme are provided in the Methods section. 

The $\Phi_{1f}$ image corresponds to the stray magnetic field generated by the current distribution in the sample. A striking feature in the $\Phi_{1f}$ image (Fig. 1c) is the strong signal observed at the top corner of the interface between the gold contact and the \CrBST{} channel which gradually evolves into a smooth gradient along the $y$ direction further away from the contact. 

To understand the current distribution in the vicinity of the contact, we performed electrostatic potential simulations of the interface between the metallic contact and \CrBST{} channel. In this model, the \CrBST{} film is represented by a resistor network \cite{sample1987reverse}, which includes elements to represent both the longitudinal and transverse resistivity with the resistor value determined by $\rho_{xx}$ and $\rho_{xy}$ measured under the same $I_b$ and $V_{BG}$ as the data in Fig. 1c. The metallic contacts are modeled by resistive elements that have a vanishing Hall effect. The result of the simulation is only sensitive to the change in Hall angle, $\theta_H$, defined through $\tan{\theta_H}=\rho_{xy}/\rho_{xx}$ across the interface (see Supplementary Information: Resistor Network Model for details). In Fig 1d, we show the simulated electrostatic potential and current density for current applied to the contact. The dramatic change in $\theta_H$ at the interface between the contact and the channel strongly distorts the electrostatic potential and current distributions, leading to a concentration of the current density at one corner of the contact. To compare modeling and data, we calculate the magnetic field corresponding to the model current distribution, and convolve it with the imaging kernel of our SQUID. The resulting model $\Phi_{1f}$ image, shown in Fig 1e, quantitatively captures the measured $\Phi_{1f}$ image. In Supplemental Fig. 2, we show that we have similar quantitative agreement between our magnetic imaging and modeling as we tune $\theta_H$ with the back gate voltage. In a separate manuscript, we examine additional contributions to $\Phi_{1f}$ arising from an interplay of bias-induced shifts in the chemical potential and corresponding changes in the magnetization at values of $I_b$ substantially exceeding those considered here \cite{ferguson2025imaging}. 
 
In Fig. 1g, we find that $\Phi_{2f}$ has a pronounced signal in the corner of the contact coinciding with the ``hot-spot" observed in the current density. The sign, spatial structure and amplitude of the $\Phi_{2f}$ signal are consistent with a small demagnetization of the sample, which we establish below is in response to current-induced heating of the electrons. Fig. 1h shows the corresponding change in sample magnetization, $\delta M_{2f}$ reconstructed from $\Phi_{2f}$. The change in magnetization is strongest at the ``hot-spot" where Figs. 1c-e indicate the current density is the highest, and we expect the most significant heating. In agreement with the current density, the hot-spot moves from the top to the bottom of the contact when the magnetization is reversed (Extended Data Fig. 1). 

Our observations bear a striking resemblance to work on dissipation in the integer quantum Hall regime in GaAs/AlGaAs heterostructures. Early work utilizing the fountain effect in a superfluid helium film to monitor the dissipation in a Hall bar device showed that power dissipation was highest in the two opposite corners of current-injecting contacts, with the corner determined by the sign of the magnetic field \cite{klass1991fountain, klass1992image}. Similarly, spatially resolved cyclotron emission studies revealed enhanced emission near the current entry and exit corners \cite{kawano1999cyclotron, komiyama2006electron}. 

Next, we focus on images of $\delta M_{2f}$ as we vary the applied current. Fig. 2a shows $\rho_{xx}$ and $\rho_{xy}$ as a function of $I_{b}$. At low values of $I_{b}$, the sample exhibits the QAHE: the Hall resistivity is quantized in units of $h/e^2$ and the longitudinal resistivity $\rho_{xx}$ is vanishing ($\rho_{xx} \approx \SI{45}{\ohm}$). As $I_b$ is increased beyond $\sim \SI{20}{\nano\ampere}$, $\rho_{xx}$ increases and quantization of $\rho_{xy}$ lost, indicating the breakdown of the QAHE. 

\begin{figure}
    \centering
    \includegraphics[width=1.0\textwidth]{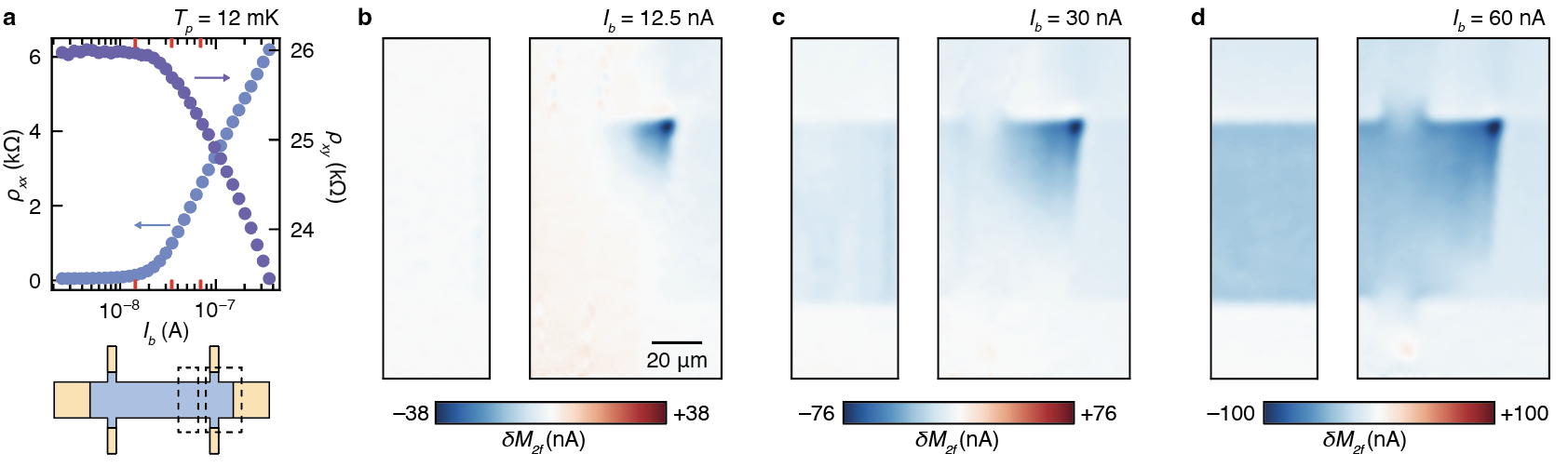}
    \caption{\textbf{QAH breakdown via bias-induced heating a,} Longitudinal ($\rho_{xx}$) and Hall ($\rho_{xy}$) resistivities versus bias current ($I_b$). Red tick marks indicate bias points for images in \textbf{b-d}, and their locations are depicted schematically at the bottom of the panel. \textbf{b,} $\delta M_{2f}$ images for $I_b = \SI{12.5}{\nano\ampere}$ below the breakdown of the QAHE. Strong signatures of heating are observed near the hot-spot, but not near the voltage probes or in the channel. \textbf{c,} $\delta M_{2f}$ images for $I_b = \SI{30}{\nano\ampere}$ at the onset of breakdown  showing signatures of heating near the voltage probes and in the channel. \textbf{d,} $\delta M_{2f}$ images at $I_b = \SI{60}{\nano\ampere}$ above the breakdown current for the QAHE with nearly uniform heating signatures away from the hot-spot.}
\end{figure}

In Figs. 2b-d we show the reconstructed $\delta M_{2f}$ at three representative values of $I_{b}$. At $I_{b} = \SI{12.5}{\nano\ampere}$ (Fig. 2b), which is below the breakdown threshold, no heating is observed in the channel (left), however signatures of heating are still evident at the contact hot-spot (right). For $I_{b} = \SI{30}{\nano\ampere}$, the breakdown of the QAHE begins. $\rho_{xx}$ is finite, and a heating signal appears both in the channel (right) as well as near the contact region (left). By the time $I_{b}$ reaches \SI{60}{\nano\ampere} (Fig. 2d), we observe a more pronounced heating signal throughout the channel. A strong correlation between the magnitude of $\delta M_{2f}$ and the electrical transport data is clear. A strong $\delta M_{2f}$ consistently appears near the corner of the contact, where the current density is so high that breakdown is induced locally, even when dissipation measured by $\rho_{xx}$ in the channel is low. Once $I_b$ becomes large enough to cause an increase in $\rho_{xx}$ as measured in transport, a corresponding $\delta M_{2f}$ is observed throughout the entire device channel.

To study the correspondence between electrical transport data and the observed $\delta M_{2f}$ in more detail, we introduce a thermal model of the system, schematically illustrated in Fig. 3a. This model consists of two subsystems: the electrons, described by a temperature $T_e$, and the lattice at a temperature $T_p$. Within our model, the electronic subsystem absorbs energy at a rate given by $P_{in} = R_{xx}I_b^2$ with $R_{xx}$ a function of $T_e$. The electrons dissipate energy absorbed from the externally applied bias to the lattice at temperature $T_{p}$ through an effective thermal resistance $R_{th}$, which represents all thermal relaxation processes between the electrons and their environment.

To characterize $R_{th}$, we apply the $3\omega$ technique, which allows us to use the temperature dependence of $R_{xx}$ to detect changes in $T_e$ induced by $I_b$ \cite{cahill1987thermal, cahill1990thermal}. When $R_{xx}$ depends on $T_e$, changes to $T_e$ induced by $I_b$ generate a third-harmonic voltage drop over the sample $V_{3f}$, due to variations in the sample temperature and resistance during the duty-cycle of the lock-in excitation. The amplitude of $V_{3f}$ depends on the temperature dependence of $R_{xx}$ and $R_{th}$ alone, allowing us to characterize $R_{th}$ via measurements of $\rho_{xx}$ and the bias and temperature dependence of $V_{3f}$. Further details of the $3\omega$ technique as well as consistency checks are presented in the Methods and Supplementary information.

In Fig. 3b, we show the $I_b$ and $T_p$ dependence of the longitudinal resistance $R_{xx}$ for our device. At each $T_p$, $R_{xx}$ is minimized and independent of bias for low values of $I_b$, indicating that $T_p = T_e$ at low bias. From this low-bias data, we determine the dependence of $R_{xx}$ on $T_e$. We then use this dependence to extract $R_{th}$ from the measured $V_{3f}$, as shown in Fig. 3c. We find that $R_{th}$ follows a power-law temperature dependence given by $R_{th} = \left(A \Sigma\right)^{-1} T^{-n}$ where $A$ is the area of the \CrBST{} channel and $\Sigma$ characterizes the strength of the energy relaxation processes in our \CrBST{} device. A fit to the temperature dependence of $R_{th}$ yields best-fit parameters $n = 4.0\pm0.1$ and $\Sigma = \SI{10 \pm 2}{\watt\meter^{-2}\kelvin^{-5}}$.

Our phenomenological model for energy relaxation in our device does not assume a specific microscopic process described by $R_{th}$. Previous work on energy relaxation at millikelvin temperatures in metals \cite{roukes1985hot, wellstood1994hot}, bulk semiconductors \cite{zhang1998non}, and GaAs heterostructures  in the integer quantum Hall regime \cite{chow1995experiments, chow1996phonon} have consistently identified power-law energy relaxation processes, $R_{th} \sim T^{-\alpha}$, with $\alpha$ between 3 and 5, depending on the material system. In these studies, $R_{th}$ is attributed to electron-phonon coupling. In \CrBST{}, direct electron-phonon mediated energy relaxation is also a plausible microscopic process which establishes a thermal link between the electrons and the environment. Additionally, given the magnetic order in \CrBST{}, energy relaxation may also have contributions from collective excitations associated with the magnetization. The $R_{th}$ that we report represents the combined effect of all energy relaxation processes.

\begin{figure}
    \centering
    \includegraphics[width=0.5\textwidth]{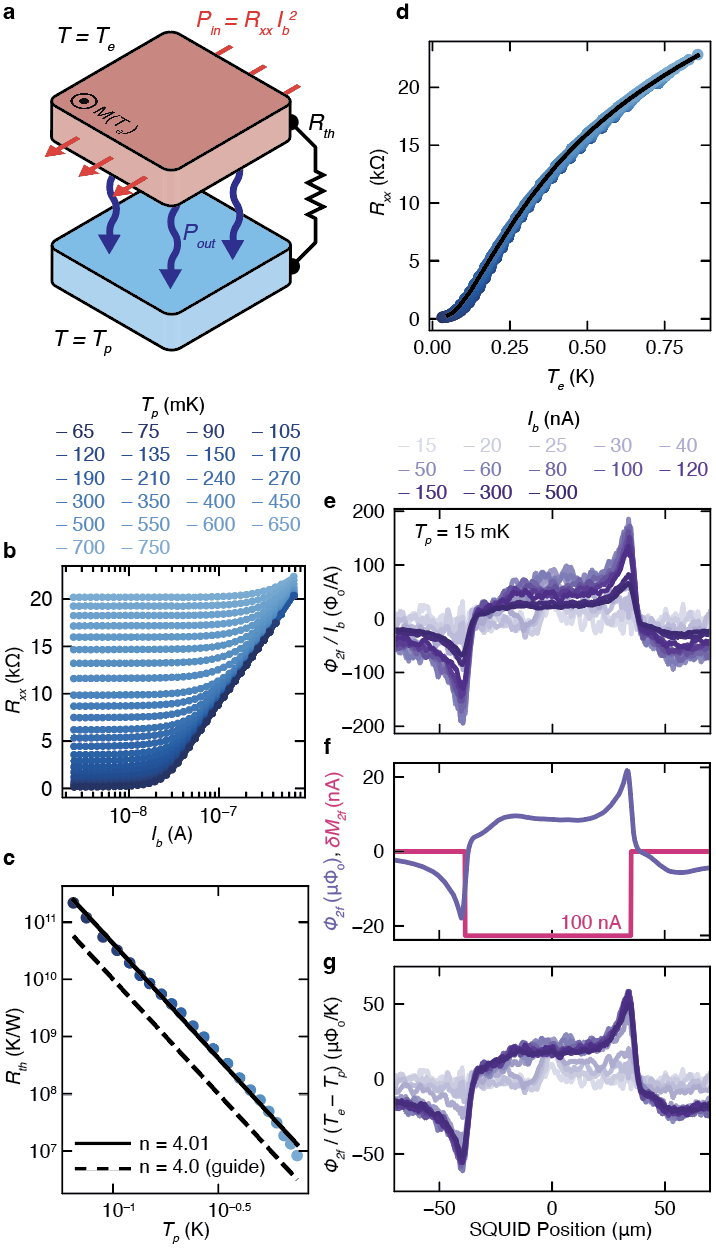}
    \caption{\textbf{Hot electrons and energy relaxation a,} Illustration of the thermal relaxation model. Electrons absorb power $P_{in}$ from the bias current. Power is dissipated to the environment through the thermal resistance $R_{th}$. \textbf{b,} Longitudinal resistance $R_{xx}$ as a function of bias current $I_b$ and lattice temperature $T_p$. \textbf{c}, Temperature dependence of the thermal relaxation strength $R_{th}$, measured via the $3\omega$ method at the same $T_p$ as in (\textbf{b}). \textbf{d,} $R_{xx}$ from \textbf{(b)} plotted versus the electron temperature $T_e$ inferred from the model in (\textbf{a}) and characterized in \textbf{b, c}. The data collapse onto a single curve, indicating that $R_{xx}$ depends solely $T_e$. The black curve shows the temperature dependence of $R_{xx}$ for low bias ($I_b$ = \SI{2.45}{\nano\ampere}) for comparison. \textbf{e,} $\Phi_{2f}$ acquired along a line over the device channel at $T_p = \SI{15}{\milli\kelvin}$ for various $I_b$. \textbf{f,} Model of a spatially uniform demagnetization of the channel $\delta M_{2f}$ of \SI{100}{\nano\ampere} is shown in blue. Convolving the $\delta M_{2f}$ signal with the SQUID's imaging kernel yields the corresponding modeled flux $\Phi_{2f}$. \textbf{g,} Data from (\textbf{e}), normalized by the electron heating amplitude $T_e – T_p$ calculated from the hot electron model. Traces with $T_e - T_p < \SI{25}{\milli\kelvin}$ are omitted for clarity.}
\end{figure}

With the temperature dependence of $R_{th}$ established, we can estimate $T_{e}$ for each combination of $T_p$ and $I_b$. For electrons and phonons initially in equilibrium at temperature $T$, $\mathrm{d}P_{out} = R_{th}(T)^{-1} \mathrm{d}T$ represents the incremental increase in the power dissipated to the environment for an incremental increase in the electron temperature $\mathrm{d}T$. In steady state, the input power to the electronic subsystem $P_{in}$ is equal to the power dissipated to the environment. Integrating $\mathrm{d}P_{out}$ from $T_p$ to $T_e$ to obtain $P_{out}$, and setting $P_{in} = P_{out}$ yields,
\begin{equation}
T_e = \left[\frac{5P_{in}}{A\Sigma} + T_p^5 \right]^{{1/5}},
\label{eq:Te_formula}
\end{equation}
where we have used $R_{th} = \left(A \Sigma\right)^{-1} T^{-4}$, based on our $3\omega$ measurements. Since $P_{in}$ depends on $T_e$ through the temperature dependence of $R_{xx}$, we solve Eq. \ref{eq:Te_formula} numerically for $T_e$ using our measurements of $R_{th}$ and $R_{xx}$. In Fig. 3d, we re-plot the data from Fig. 3b as a function of the resulting $T_{e}$ calculated with our heating model. The data collapse onto a single curve, independent of the specific values of $T_p$ and $I_b$ at which $R_{xx}$ was measured. Additionally, we highlight in Fig. 3d the values of $R_{xx}$ measured at the lowest bias limit where $T_e = T_p$. The re-scaled data for each combination of $T_p$ and $I_b$ agree closely with this low-bias curve. The same analysis also allows us to collapse the $R_{xy}$ data onto a single curve independent of $T_p$ and $I_b$ (Supplemental Information: Consistency checks for the $3\omega$ method.).

The collapse of the electronic transport coefficients onto a single curve indicates that nearly all changes in the sample's resistivity with both $I_b$ and $T_p$ can be attributed to changes in the electron temperature $T_e$. Consequently, the breakdown of the QAHE in our device is solely governed by $T_e$. While changing $T_p$, $I_b$, or both may induce breakdown, these changes ultimately generate a corresponding change in $T_e$, restoring balance between $P_{in}$ and $P_{out}$.

With the energy absorption and relaxation processes characterized, we now turn to examining the local magnetic signatures of heating. In Fig. 3e, we plot $\Phi_{2f}$ normalized by $I_b$ acquired along a line between the voltage probes in our Hall bar at various values of $I_b$. At the lowest values of $I_b$, there is minimal spatial structure in $\Phi_{2f}$, as dissipation in the channel is low and changes in $T_e$ in response to $I_b$ are small. Fig. 3f shows a simulated magnetic flux profile for a uniform sample demagnetization across the channel width. At higher bias currents, the spatial profile of $\Phi_{2f}$ in Fig. 3e closely resembles this model for a uniform demagnetization. Fig. 3e reveals that the magnitude of $\Phi_{2f} / I_b$ obeys a sub-linear dependence on $I_b$: traces at the highest values of $I_{b}$ exhibit a smaller peak-to-peak amplitude than those at intermediate values of $I_b$. An interpretation of $\Phi_{2f}$ in terms of a response proportional to the dissipated power, $\Phi_{2f} \sim P_{in} \sim R_{xx}I_{b}^2$, is incompatible with the data.

To understand the scaling of $\Phi_{2f}$, we re-plot the traces from Fig. 3e in Fig. 3g, now normalizing $\Phi_{2f}$ by the temperature difference between the electrons and phonons, $T_{e} - T_{p}$, as predicted by our heating model. This separates the curves into two classes: curves for $I_b$ below $\sim \SI{30}{\nano\ampere}$ exhibit little spatial structure, whereas above this level they collapse onto a single profile. Notably, $I_b \sim \SI{30}{\nano\ampere}$ is also the value beyond which $R_{xx}$ in Fig. 3 rises significantly. Taken together, these observations indicate that $\Phi_{2f}$ is proportional to changes in $T_e$. By tracking changes in $T_e$ through corresponding changes in the magnetization, we spatially resolve regions of the sample where $T_e$ is driven out of equilibrium by the bias current with micron-scale spatial resolution. We repeated this measurement and analysis at several values of $T_p$, and present the results in Extended Data Fig. 2. As $T_p$ increases, the thermal coupling between the electrons and lattice becomes stronger and the bias current required to drive $T_e$ above $T_p$ increases. Similarly, increasing $T_p$ also changes the heating signatures in the vicinity of the hot spot. In Extended Data Fig. 3, we show images of $\Phi_{DC}$, $\Phi_{1f}$, $\Phi_{2f}$, and reconstructed $\delta M_{2f}$ acquired near the contact interface, with $T_p = \SI{750}{\milli\kelvin}$ and same bias current, $I_{b} = \SI{150}{\nano\ampere}$, as the data in Fig. 1. Although signatures of heating are dramatically weakened by raising $T_p$, we still observe heating effects localized at the hot spot. 

The temperature differences generated between the electrons and the lattice during breakdown are substantial. Our measurements indicate that even for modest bias currents of $\sim \SI{100}{\nano \ampere}$, a temperature difference of several hundred mK is generated between the electronic and lattice subsystems. The amplitude of the temperature differences generated between $T_e$ and $T_p$ for the range of bias currents and $T_p$ explored in this work is presented in Extended Data Figure 4. The heating effect is even more pronounced near the contacts, where the current density at the hot spot is high regardless of the total source drain bias. In this region, substantial heating occurs even when $\rho_{xx}$ near the voltage probes approaches zero. The large electron and phonon temperature differences may play a role in a number of phenomena reported in transport experiments, including electronic switching of magnetically doped topological insulators driven by source-drain bias currents several orders of magnitude larger than those used here \cite{yuan2024electrical, zimmermann2024current}.


To understand the details of energy dissipation near the hot spot, we extend our resistor network model to account for heating effects. Using the electrostatic potential distribution obtained for $T_e = T_p$, we calculate the local power dissipation per unit area, $p_{in}$ within the device. By identifying $P_{in}/A$ with $p_{in}$ in Eq. \ref{eq:Te_formula}, we can use $\Sigma$ extracted from the $3\omega$ method to determine the local $T_e$ distribution. Using this temperature distribution, we update the local conductivity values and re-calculate the electrostatic potential distribution, $p_{in}$ and $T_e$. After a few iterations, these calculations converge to a self-consistent temperature distribution for a given combination of $I_b$ and $T_p$, from which all other properties of the device may be calculated. In this model, we neglect the effects of thermal transport within the electronic subsystem and consider only the energy transport processes described by $R_{th}$ which relax energy out of the electronic subsystem to the environment. We discuss the effects of thermal transport through the electrons to the metallic contacts below. Further details of the self-consistent calculations are presented in Supplementary Information: Resistor network model.

In Fig. 4a, we show the electron heating ($T_e - T_p$) for the experimental conditions ($T_p < \SI{50}{\milli\kelvin}$, $I_b = \SI{12.5}{\nano\ampere}$) and field of view as in Fig. 2b. Substantial heating is found in the vicinity of the hot-spot, where the current density is highest. Due to the high conductivity of the metallic contact, $p_{in}$ is several orders of magnitude smaller than in the semiconducting channel, and heating in the contacts is negligible in our model. Away from the contact area, we do not find substantial heating for $I_b = \SI{12.5}{\nano\ampere}$. In Fig. 4b, we show results for $I_b = \SI{60}{\nano\ampere}$, a bias current above the threshold for breakdown of the QAHE in our device (see Fig. 2). In contrast to $I_b = \SI{12.5}{\nano\ampere}$, a uniform temperature difference $T_e - T_p \approx \SI{75}{\milli\kelvin}$ develops in the device channel. 

A quantitative comparison of this calculation to images of $\Phi_{2f}$ requires the temperature dependence of the sample magnetization $M(T)$. Given that typical values of $T_e$ are small compared to the Curie temperature of $\sim \SI{20}{\kelvin}$ under all conditions investigated in this work \cite{chang2023review}, we adopt the  approximation $M(T) \approx M_o - \alpha T$, appropriate for superparamagnetic thin-films far below the Curie temperature \cite{qiu1989thermal, lachman2015visualization, lachman2017observation}. For the $\Phi_{2f}$ acquired over the channel in Fig. 3e, we find that $\alpha = \SI{300}{\nano\ampere/\kelvin}$ provides good agreement between the measured $\Phi_{2f}$ data and $T_e$ predicted by the model. Compared to the static magnetization of the sample, $M_o$, heating-induced changes in the magnetization are small. From $\Phi_{DC}$ measured scanning the SQUID over the channel, we estimate that $M_o \approx \SI{100}{\micro\ampere} \gg \alpha \SI{1}{\kelvin} \approx \SI{300}{\nano\ampere}$, where a temperature change of $\sim \SI{1}{\kelvin}$ represents a rough upper bound for the heating effects in our sample. This estimate of $M_o \approx \SI{100}{\micro\ampere}$ gives a magnetization per Cr-doped layer which is consistent with previous measurements in similar \CrBST{} samples \cite{lachman2015visualization}.

In Fig. 4c and d, we show simulated $\Phi_{2f,m}$ images obtained by calculating $\delta M_{2f,m}$ using $\alpha$ and the local $T_e$, calculating the corresponding stray magnetic field and convolving  with the imaging kernel of our SQUID. For comparison, Fig. 4e and f show the corresponding $\Phi_{2f}$ images, acquired under experimental conditions matching the model, with additional comparisons for other values of $I_b$ presented in Extended Data Figure 5. Despite its simplicity, the model shows good agreement with the data: strong $\Phi_{2f}$ signals consistently appear at the hot-spots indicating large temperature differences between the electrons and phonons. Away from the contacts, $\Phi_{2f}$ signals are associated with the onset of dissipation in the channel and the breakdown of the QAHE detected in electrical transport measurements. 

In the simulation, we assume that the thermal conduction through the electronic system is negligible compared to $R_{th}$. Although this assumption yields good quantitative agreement with the $\Phi_{2f}$ data away from the contacts, the model systematically over-estimates the $\Phi_{2f}$ signal at the hot spot. Both our data and simulations indicate that the metallic contacts remain near $T_p$, leading to pronounced temperature gradients near the contacts. A plausible explanation for the discrepancy is the electronic thermal conduction providing cooling from the metallic contact at the semiconductor-metal interface. An estimate of the electronic contribution to the thermal conductance indicates that this cooling pathway is important only within a few micrometers of the semiconductor-metal interface (See Supplementary Section: Estimate of the Electronic Thermal Conductance). Due to the vanishing $\sigma_{xx}$ at low temperatures, energy absorbed by the electrons in the channel cannot be transported to the contacts before it is relaxed directly through $R_{th}$. 


Heating effects have previously been considered in the context of the breakdown of the integer quantum Hall effect (IQHE). In high mobility samples, the breakdown of the IQHE was found to be non-local \cite{komiyama1996evidence, kaya1998spatial}, with hot electrons diffusing over hundreds of micrometers. In contrast, in our measurements, as well as in transport measurements on similar \CrBST{} samples \cite{rosen2022measured}, the breakdown phenomenology may be understood without introducing non-local behavior. This discrepancy may qualitatively be understood in terms heavy doping required to realize the QAHE in MTIs, which severely limits the carrier mobility. The QAHE has been realized in a variety of material platforms including \BST{} doped with different magnetic elements, stoichiometric magnetic topological insulators \cite{deng2020quantum}, and van der Waals heterostructures \cite{serlin2020intrinsic, li2021quantum, han2024correlated, han2024large, zhao2024realization}. We anticipate that differences in the carrier mean-free-path across material systems will qualitatively alter the nature of energy absorption and relaxation in the QAH regime. In particular, measurements on higher mobility QAH systems may identify non-local heating and breakdown behavior similar to those detected in the IQHE in GaAs. 


\begin{figure}
    \centering
    \includegraphics[width=0.5\textwidth]{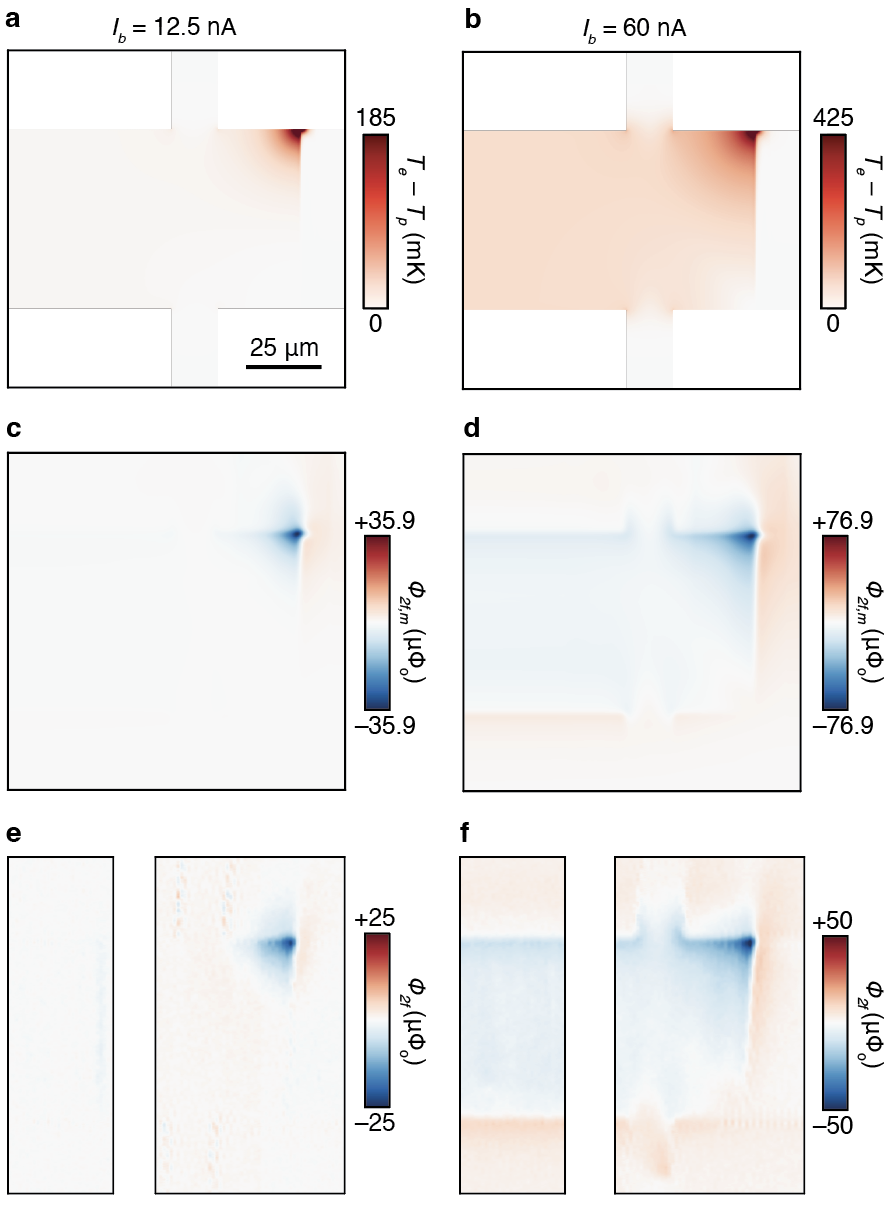}
    \caption{\textbf{Imaging hot electrons a,} Spatial profile of the heating effect, $T_e – T_p$, calculated self-consistently with the current and electrostatic potential distributions, using the thermal relaxation strength extracted from the $3\omega$ technique. Model $I_b$ and $T_p$ are chosen to match the experimental conditions in Fig. 2b, with $I_b$ below the breakdown current for the QAHE. \textbf{b,} Same as \textbf{(a),} with $I_b$ and $T_p$ adjusted to the experimental conditions of Fig. 2d, with $I_b$ above the breakdown current for the QAHE. \textbf{c,} Model $\Phi_{2f,m}$ calculated using the temperature profile in (\textbf{a}), assuming a temperature dependence of the magnetization with $\alpha =\SI{300}{\nano\ampere/\kelvin}$ \textbf{d,} Same as (\textbf{c}), but using the temperature profile in \textbf{(b)}. \textbf{e, f} Experimental $\Phi_{2f}$ used to reconstruct $\delta M_{2f}$ shown in (\textbf{e}) Fig. 2b and (\textbf{f}) in Fig. 2d.}
\end{figure}

Although energy relaxation processes impose fundamental limitations on the performance of any device which utilizes the novel electronic behavior exhibited by quantum materials, direct experimental access to this information remains scarce \cite{marguerite2019imaging, jalabert2023thermalization}. We have developed a method of experimentally monitoring power dissipation in \CrBST{} samples using a scanning SQUID microscope. By combining this approach with transport measurements of the energy relaxation rate, we identified the central role that hot electrons play in regulating the breakdown of the QAHE. We find that breakdown is well described by a local picture of energy relaxation. As the breakdown of the QAHE is investigated in new materials systems, particularly those with higher mobility, our approach will guide the characterization of breakdown as the QAHE is pushed towards higher temperatures and current densities.

\subsection*{Acknowledgements}
We acknowledge helpful discussions at the early stages of this work with Erik Henriksen, Brad Ramshaw, Ga{\"e}l Grissonnache, and David Goldhaber-Gordon. Work at Cornell University was primarily supported by the U.S. Department of Energy, Office of Basic Energy Sciences, Division of Materials Sciences and Engineering, under award DE-SC0015947 with additional support from a New Frontier Grant awarded by Cornell University's College of Arts \& Sciences. Sample synthesis and fabrication at Penn State was supported by the Penn State 2DCC-MIP under NSF Grant Nos. DMR-1539916 and DMR-2039351.


\bibliography{references}

\setcounter{figure}{0}
\renewcommand{\thefigure}{ED\arabic{figure}}

\section*{Methods} 
\subsection*{Sample growth and sample fabrication}
We used a VEECO 620 molecular beam epitaxy (MBE) system to grow heterostructures comprised of 3 quintuple layer (QL) \CrBST - 5QL \BST - 3QL \CrBST ~on \ce{SrTiO3}  (111) substrates (MTI Corporation). The Cr composition is nominal (based on past calibrations). The \ce{SrTiO3} substrates were cleaned using deionized water at \si{90}{\celsius} for 1.5 hours and thermally annealed at \si{985}{\celsius} for 3 hours in a tube furnace with flowing oxygen gas. The substrate was out-gassed under vacuum at \si{630}{\celsius} for 1 hour and then cooled down to \si{340}{\celsius} for the heterostructure growth. When the temperature of substrate was stable at \si{340}{\celsius}, high-purity Cr (5N), Bi (5N), Sb (6N), and Te (6N) were evaporated from Knudsen effusion cells to form the heterostructure. The desired beam equivalent pressure (BEP) fluxes of each element and the growth rate were precisely controlled by the cell temperatures. The BEP flux ratio of Te/(Bi + Sb) was kept higher than 10 to prevent Te deficiency. The BEP flux ratio of Sb/Bi was kept around 2 to tune the chemical potential of the heterostructure close to the charge neutrality point. The heterostructure growth rate was ~0.25 QL/min, and the pressure of the MBE chamber was maintained at $2 \times 10^{-10}$ mbar during the growth.

After the growth, the heterostructures were fabricated into a \SI{200}{\micro \meter} $\times$ \SI{75}{\micro \meter} Hall bar and a two-terminal sample using photolithography. The shape of the samples was defined by Argon plasma etching. After etching, 10 nm Cr/60 nm Au were deposited outside the active area of the Hall bar to make electrical contact. The top gate was fabricated by depositing a 40 nm \ce{Al2O3} layer by atomic layer deposition across the entire sample and evaporating a 10 nm Ti/60 nm Au layer patterned by optical lithography.

\subsection*{Electronic transport characterization}
Electrical connection to the samples were made via thermocoax lines in a cryogen-free dilution refrigerator with a base temperature of $\sim\SI{15}{\milli\kelvin}$ at the mixing chamber plate. Samples are mounted on a high thermal conductivity copper cold finger in the bore of a 6T-1T-1T vector magnet. We use the reference output of a Stanford Research Systems SR830 lock-in amplifier to excite one current contact of our Hall bar with a sinusoidal excitation at a frequency of \SI{140.5}{\hertz} while leaving the other current contact grounded. The longitudinal and Hall voltages are amplified by Stanford Research Systems SR540 preamplifiers before being demodulated by two lock-in amplifiers. The carrier density in the sample was tuned using the sample back gate formed by the \STO{} substrate. During electrical transport characterization and current density imaging, the top gate was grounded.

\subsection*{Measurement of \Rth{} with the $3 \omega$ method}
To measure the thermal resistance characterizing energy relaxation from the electrons to the environment, we use the temperature-dependent resistance of the sample as a thermometer. We source a current $I = I_o \sin{\omega t}$ through the sample, generating heat in the sample at both DC and at $2\omega$. The input power $P_{in}$ generates a temperature difference between the electrons at temperature $T_e$ and the phonon bath at temperature $T_p$. At the low frequencies used in our experiments, this temperature difference appears in-phase with the the heat flow into the sample and \Rth{} may be defined,
\begin{equation*}
    R_{th} = \frac{\Delta T}{P_{in}},
\end{equation*}
Where $\Delta T = T_{e} - T_{p}$, is the difference between the electron and phonon temperatures. The change in $T_e$ in response to the input power causes the sample resistance to change. For small $\Delta T$, $R_{xx}(T)$ may be expressed,
\begin{equation*}
    R_{xx}(T_e) = R_{xx}(T_p) + \frac{dR}{dT} \Delta T = R_{xx}(T_p) + \frac{dR}{dT} R_{th}(T_p) R_{xx}(T_p) I^{2},
\end{equation*}
Where we have used $P_{in} = R_{xx}(T_p) I^{2}$. Next we introduce the time-dependence of the bias current, $I = I_o \sin{\omega t}$, and write the voltage drop over the sample using $V(t) = I(t) R(T)$,
\begin{equation*}
    V_{1\omega}(t) = I_o R(T_p) \left[ 1 + \frac{3}{4} \frac{dR}{dT} I_{o}^{2} R_{th} \right] \sin{\omega t}
\end{equation*}
\begin{equation*}
    V_{3\omega}(t) = -I_o R(T_p) \left[ \frac{1}{4} \frac{dR}{dT} I_{o}^{2} R_{th} \right] \sin{3\omega t}
\end{equation*}
where $V_{1\omega}(t)$ and $V_{3\omega}(t)$ are the first and third harmonic components of the voltage drop over the device. In a lock-in measurement, the the amplitude of the third-harmonic voltage may then be used to extract $R_{th}$,
\begin{equation*}
    R_{th} = \frac{4 V_{3\omega}}{I_{o}^{3} R(T_p) \alpha},
\end{equation*}
where $\alpha = dR/dT$. We use $R_{xx}(T_p)$ measured with $I_{o} = \SI{2}{\nano\ampere}$, where bias-induced heating is negligible, to calculate $dR/dT$. We then extract the slope of $V_{3\omega}$ vs. $I_{o}^{3}$ in the low-bias ($I_{o} < \SI{10}{\nano \ampere}$) limit at a range of lattice temperatures $T_p$. Using $R_{xx}(T_p)$ and its derivative, we calculate $R_{th}$ from the extracted slope of the $V_{3\omega}$ signal. At the lowest values of $T_p$, where both $R_{xx}$ and $dR/dT$ vanish, this analysis becomes unreliable. We therefore extract $R_{th}$ only for $T_p > \SI{60}{\milli\kelvin}$. $R_{th}$ may also be extracted from the $V_{1\omega}$ signal after subtracting the Ohmic contribution, comparison between the two analysis methods is described in Supplementary Section: Consistency checks for $3\omega$ technique. 

\subsection*{Extraction of $T_{e}$ from transport}
To calculate the steady-state temperature difference $T_e - T_p$ between the electrons and the phonons, we start with the condition of energy balance, which dictates $P_{in} = P_{out}$ once the system reaches steady state \cite{roukes1985hot, wellstood1994hot}. We use the resistance $R_{xx}$ and RMS bias current $I_b$ determined by lock-in measurements to estimate the input power $P_{in} = I_b^2R_{xx}$. The power dissipated to the environment is related to $R_{th}$ by $dP_{out} = \left( 1 / R_{th} \right)dT$. We integrate $dP_{out}$ from $T_p$ to $T_e$ to obtain $P_{out}$ and set $P_{out} = P_{in}$,
\begin{equation*}
     P_{in} = R_{xx}I_b^2 = \int_{T_p}^{T_e} \frac{dT}{R_{th}(T)}.
    \label{eq:integrals}
\end{equation*}
Evaluating the integral yields a simple relation between $P_{in}$, $T_{e}$ and $T_{p}$,
\begin{equation}
     P_{in}
    = \frac{A \Sigma}{5}\left[T_{e}^{5} - T_{p}^{5}\right],
    \label{eq:Te}
\end{equation}
Where we have used $R_{th} = (A\Sigma)^{-1} T^{-4}$, with $A\Sigma = \SI{4.6e6}{\kelvin^5\per\watt}$ extracted from $R_{th}$ measurements performed with the $3 \omega$ method to arrive at the analytical expression above. Here $A = \SI{75}{\micro\meter} \times \SI{300}{\micro\meter}$ is the area of the \CrBST{} channel and $\Sigma \approx \SI{10}{\watt / \meter^2 \kelvin^5}$ is the intrinsic thermal relaxation strength of our \CrBST{} heterostructure. In practice, $R_{xx}(I)$ is dependent on the RMS bias current $I_b$ we use. We expect our estimate of $P_{in}$ using a lock-in method to determine $R_{xx}$ to be reasonably accurate as long as the first-harmonic response of the sample dominates the time-dependent voltage drop over the sample for all $I_b$. We have verified that the first harmonic response of our sample is at least an order of magnitude larger than the second and third harmonic responses, indicating that this approximation is indeed valid for our sample. For samples where $R_{xx}(I)$ is monotonically increasing with $I$, our estimate of $P_{in}$ will systematically over-estimate the true power dissipated the sample, due to the lock-in amplifier performing a weighted average of the sample resistance which is more heavily weighted towards values at higher currents, where the resistance is higher. Once the temperature dependence of $R_{th}$ is known, the expression Eq.\ref{eq:Te} may be rearranged to solve for $T_{e}$ for a given $T_p$ and $P_{in}$,
\begin{equation*}
T_e = \left[\frac{5P_{in}}{A\Sigma} + T_p^5 \right]^{{1/5}}.
\end{equation*}

\subsection*{Scanning SQUID microscopy}
Unless otherwise indicated, magnetic imaging was performed with a sample cooled below \SI{50}{\milli\kelvin} in a cryogen free dilution refrigerator, described elsewhere \cite{low2021scanning}. The scanning SQUID sensor has the same gradiometric design as described in Ref. \cite{huber2008gradiometric} with a \SI{1.5}{\micro\meter} inner-diameter pickup loop. The SQUID is coupled to a SQUID-array amplifier mounted on the mixing chamber plate of the dilution refrigerator. We use a home-built piezoelectric scanner to scan the SQUID $\sim$\SI{1}{\micro\meter} above the sample surface. To measure the response of the sample to a bias current, we excite one contact of the sample with a sinusoidal excitation at a frequency of \SI{140}{\hertz} from the reference output of a Stanford Research Systems (SRS) SR830 lock-in amplifier. We ground the other current contact and use the four voltage probes to record the longitudinal and Hall potential drops over the sample using a pair of SRS SR540 preamplifiers. The SQUID signal and the output of the voltage preamplifiers are demodulated by SR830 lock-in amplifiers. We demodulate the SQUID signal at both the first and second harmonic of the excitation frequency in order to measure $\Phi_{1f}$ and $\Phi_{2f}$ respectively. In this work, the first harmonic signals, including $\Phi_{1f}$, are the in-phase component of the demodulated signal. In a lock-in measurement, the second harmonic response in the sample due to heating effects appears \SI{90}{\degree} out-of-phase with the reference signal. We therefore present the quadrature component of the demodulated lock-in signal for the $\Phi_{2f}$ data reported here. Care was taken to measure at a sufficiently low frequency such that out-of-phase components (the quadrature signal for the $\Phi_{1f}$ data and the in-phase signal for the $\Phi_{2f}$ data) were minimized.

\subsection*{Current and magnetization reconstruction}
Given that the sample thickness is more than an order of magnitude smaller than both the SQUID pickup loop radius and scan height, we treat the current density and the magnetization as two-dimensional in our analysis. The magnetic flux  $\Phi(x, y)$ at lateral position $x,y$ at height $z$ above the sample detected by the SQUID is then given by the convolution of the SQUID point spread function, $K_{\textrm{PSF}}$, and the appropriate Biot-Savart kernel, $K_{\textrm{BS}}$,
\begin{equation}
    \Phi(x, y) = K_{\textrm{PSF}}(x, y) \ast K_{\textrm{BS}}(x, y) \ast g(x, y).
    \label{eq:forward_problem}
\end{equation}
Where $\ast$ denotes a convolution defined by,
\begin{equation}
    f(x, y) \ast h(x, y) =
    \int dx' dy' f(x', y') h(x' - x, y' -y).
\end{equation}
The scalar function $g(x,y)$ may be interpreted as the magnetic dipole density, when reconstructing the sample magnetization. When flux is coupled into the SQUID by a transport current distribution, $g(x,y)$ may alternatively be identified as the current stream function, which is related to the two-dimensional current density by,
\begin{equation}
    \vec{j}(x, y) = \nabla \times [g(x, y) \hat{z}].
\end{equation}
In two dimensions, $K_{\textrm{BS}}$ is given by,
\begin{equation}
    K_{\textrm{BS}} = 
    \frac{\mu_o}{2\pi} 
    \frac{2z^2 - x^2 - y^2}{(x^2 + y^2 + z^2)^{5/2}}.
\end{equation}

We extract $K_{\textrm{PSF}}$ shown in from images of superconducting vortices acquired using a nominally identical SQUID. Reconstruction of $g(x, y)$ from a measured image $\Phi(x, y)$ which includes experimental noise is a deconvolution problem. Deconvolution problems require regularization to avoid the amplification of high spatial frequency noise in the final solution. Here we combine $K_{\textrm{PSF}}$ and $K_{BS}$ into a single linear operator $M$ such that eq. \ref{eq:forward_problem} can be written as $\Phi = Mg$, where now $g$ is a vector with length $n$ equal to the number of pixels in an image and $M$ is a $n \times n$ matrix. We chose a regularization operator $\Gamma$ that penalizes solutions that include high-frequency noise, with a regularization strength $\sigma$. We seek the $g^\ast$ that optimizes,
\begin{equation}
  g^\ast = \mathrm{min}_g \left[ \frac{1}{2}||Mg - \phi||^2 + \sigma^2 ||\Gamma g||^2 \right].
\end{equation}
$g^\ast$  can be found by solving the linear equation,
\begin{equation}
    (M^T M + 2 \sigma^2\Gamma^T \Gamma) g = M^T \phi.
\end{equation}
$M^T$ and $\Gamma^T$ are the pseudo-inverse of $M$ and $\Gamma$, respectively. In practice, we do not directly calculate the elements of $M$, but instead calculate the convolution $Mg$ using Fast Fourier transforms. We use the Wiener filter to approximate $M^T$ and choose the discrete Laplace operator as our regularization operator $\Gamma$. For the one-dimensional line cut data, we utilize the same methods described above in one dimension. In this case, the SQUID point spread function and Biot-Savart kernel are integrated along one axis to form a 1D point spread function. 

\begin{figure}
    \centering
    \includegraphics[width=1.0\textwidth]{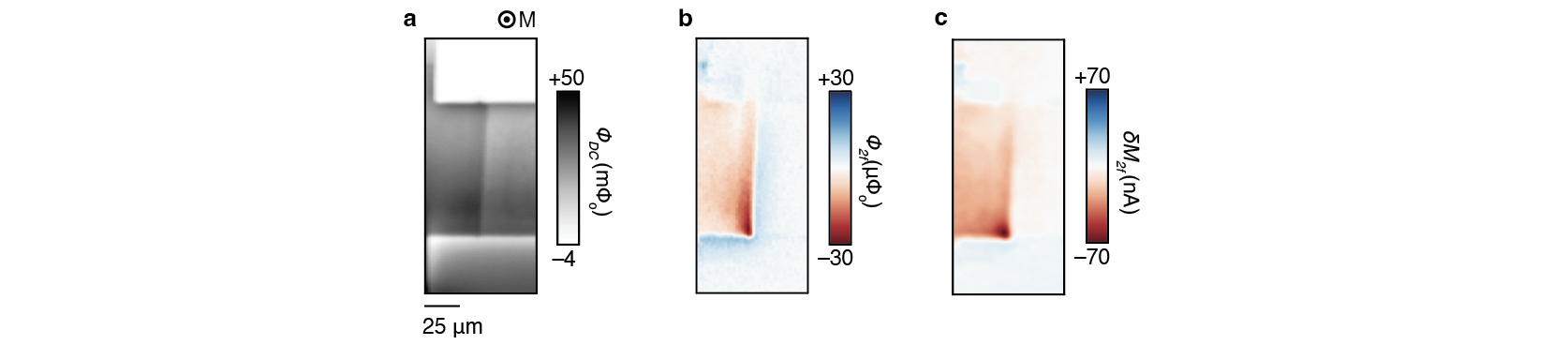}
    \caption{\textbf{Hot spot location changes for reversed magnetization a,} DC flux $\Phi_{DC}$, coupled into the SQUID pickup loop while imaging near the interface between the metal contact and semiconducting channel. The sample magnetization is reversed relative to the configuration in the main text (magnetized out-of-the plane). \textbf{b,} Second harmonic SQUID signal, $\Phi_{2f}$ recorded near the contact with the magnetization opposite to that of Fig. 1. The bias current $I_b$ is \SI{60}{\nano\ampere}. \textbf{c,} Change in the sample magnetization $\delta M_{2f}$ from bias-induced heating, reconstructed from (\textbf{b}).} 
\end{figure}

\begin{figure}
    \centering
    \includegraphics[width=1.0\textwidth]{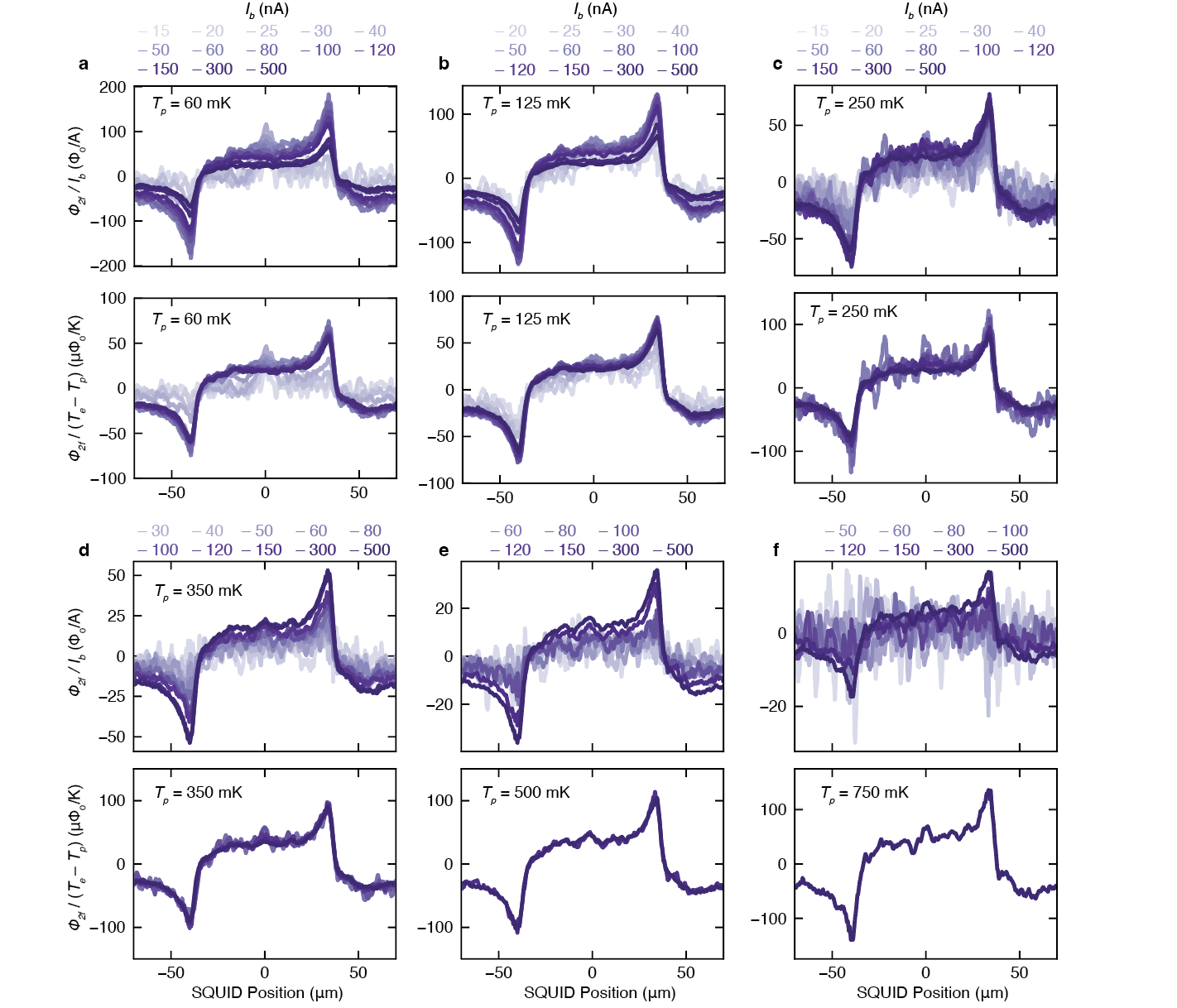}
    \caption{\textbf{Magnetic signatures of heating at higher $T_p$ a,} (top) $\Phi_{2f}$ signal acquired scanning over the width of the channel as in Fig. 4b, at a selection of source-drain bias currents $I_b$. Traces are normalized by $I_b$. $T_p = \SI{60}{\milli\kelvin}$. (bottom) $\Phi_{2f}$ signal normalized by temperature difference $T_e - T_p$ predicted from the heating model described in the main text. Traces with $T_e - T_p < \SI{25}{\milli\kelvin}$ are omitted for clarity. \textbf{b,} same as (\textbf{a}), with $T_p = \SI{125}{\milli\kelvin}$. \textbf{c,} same as (\textbf{a}), with $T_p = \SI{250}{\milli\kelvin}$. \textbf{d,} same as (\textbf{a}), with $T_p = \SI{350}{\milli\kelvin}$. \textbf{e,} same as (\textbf{a}), with $T_p = \SI{500}{\milli\kelvin}$. \textbf{f,} same as (\textbf{a}), with $T_p = \SI{750}{\milli\kelvin}$.} 
\end{figure}

\begin{figure}
    \centering
    \includegraphics[width=1.0\textwidth]{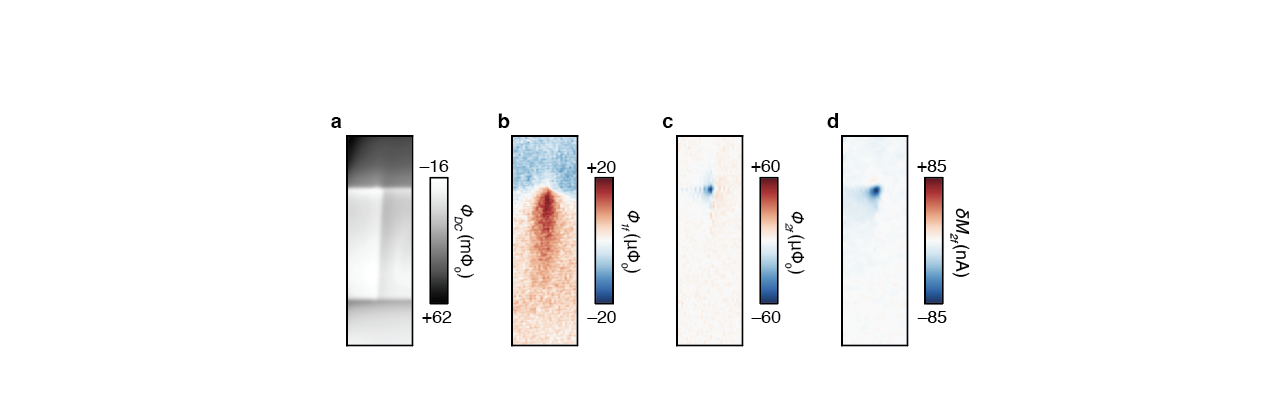}
    \caption{\textbf{Imaging bias-induced heating at elevated $T_p$ a,} DC flux, $\Phi_{DC}$, coupled into the SQUID pickup loop while imaging the interface between the Au contact and the channel. The sample is magnetized into the plane. \textbf{b,} First harmonic SQUID signal $\Phi_{1f}$ co-recorded with the image in (\textbf{a}). The bias current $I_b$ is \SI{150}{\nano\ampere}. \textbf{c,} Second harmonic SQUID signal, $\Phi_{2f}$ co-recorded with the images \textbf{a,b}. At elevated $T_p$, $\Phi_{2f}$, is suppressed compared to the $T_p = \SI{15}{\milli\kelvin}$ case shown in Fig. 1d in the main text. \textbf{d,} Change in the sample magnetization $\delta M_{2f}$ reconstructed from (\textbf{c}). Data presented here was acquired with $T_p = \SI{750}{\milli\kelvin}$. The amplitude and spatial extent of $\delta M_{2f}$ is reduced compared to the $T_p < \SI{50}{\milli\kelvin}$ case shown in Fig. 1f in the main text.
    } 
\end{figure}

\begin{figure}
    \centering
    \includegraphics[width=0.5\textwidth]{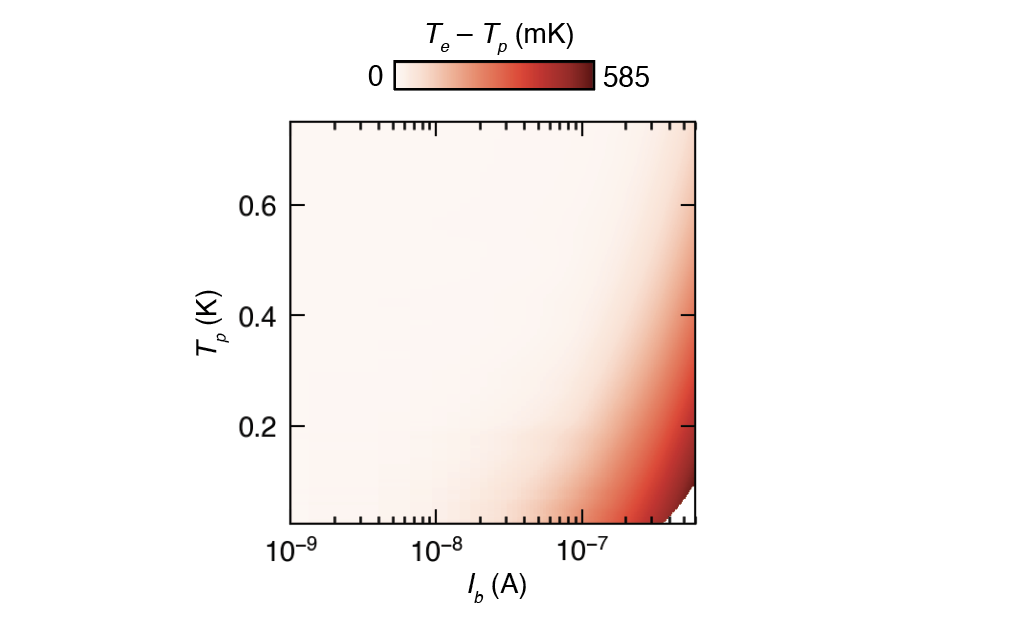}
    \caption{\textbf{$I_b$ and $T_p$ dependence of heating effect} Amplitude of the heating effect $T_e - T_p$, calculated within the hot-electron model as a function of the bias current $I_b$ and the lattice temperature $T_p$.
    } 
\end{figure}

\begin{figure}
    \centering
    \includegraphics[width=0.5\textwidth]{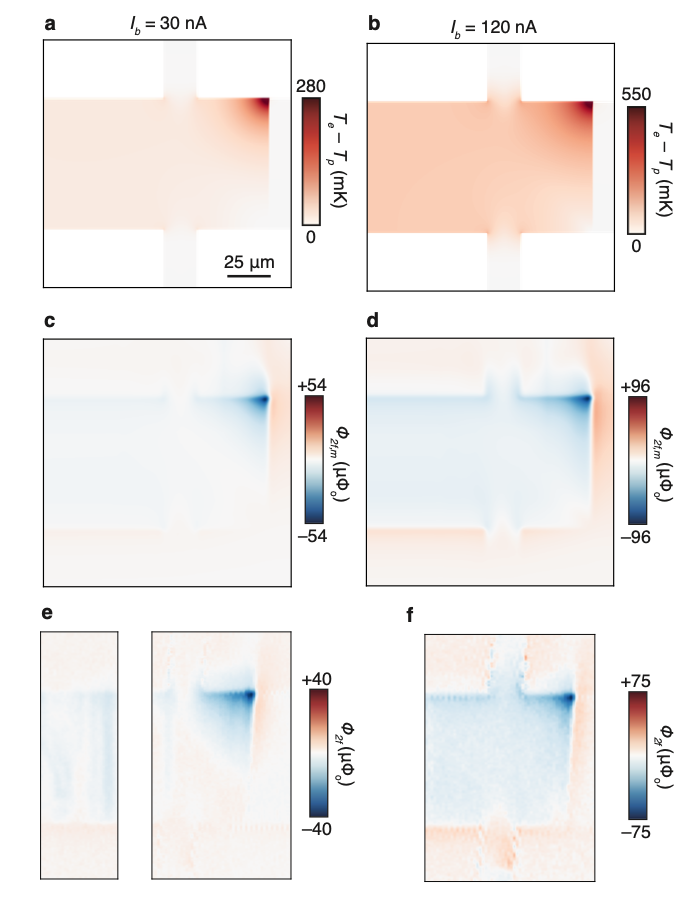}
    \caption{\textbf{Comparison of the heating model at different $I_b$ a,} Profile of the temperature difference between the electrons and phonons, calculated using the heating model for a bias current $I_b = \SI{30}{\nano\ampere}$. Model parameters are chosen to match the experimental conditions presented in Fig. 2c. \textbf{b,} Same as (\textbf{a}) for $I_b = \SI{120}{\nano\ampere}$. Model parameters are chosen to match the experimental conditions presented in Fig. 1e. Image of the model second harmonic SQUID signal $\Phi_{2f,m}$, calculated using the temperature distribution in (\textbf{a}) and assuming a magnetization temperature dependence of $\alpha =\SI{300}{\nano\ampere/\kelvin}$. \textbf{b,} Same as (\textbf{c}), calculated using the temperature profile in (\textbf{b}). \textbf{e,} Experimental $\Phi_{2f}$ measured for $I_b = \SI{30}{\nano\ampere}$. Data in (\textbf{e}) were used to reconstruct $\delta M_{2f}$ in Fig. 2c. \textbf{f,} Same as (\textbf{e}), measured with $I_b = \SI{120}{\nano\ampere}$. Data in (\textbf{f}) reproduced from Fig. 1g.
    } 
\end{figure}

\clearpage
\setcounter{page}{1}
\setcounter{figure}{0}
\renewcommand{\figurename}{Fig.}
\renewcommand{\thefigure}{S\arabic{figure}}
\renewcommand{\thesection}{S\arabic{section}}
\renewcommand{\thesubsection}{Supplementary Section \arabic{subsection}:}

\section*{Supplementary Information}

\subsection{Fitting different models for the conductivity}
To compare the electronic transport behavior of our device to previous work on the breakdown of the QAHE, we fit the conductivity of our sample to several different models for hopping transport \cite{fox2018part}. In Fig. S1 we show fits of the temperature dependence  of $\sigma_{xx}$ to three different models, with the devices tuned into the QAH regime with the back gate. In our device, the temperature dependence of $\sigma_{xx}$ levels off at the lowest temperatures measured. To make a fair comparison between the three models, we perform each fit over the same temperature range. By adjusting the temperature interval over which the fit is performed, good agreement between the any of the models and the data can be obtained. In Fig. S1a, we show a fit to a model for thermally activated conduction with $\sigma_{xx} = \sigma_o e^{-T_o/T}$. In Fig. S1b, we show a fit to a model for variable range hopping, with $\sigma_{xx} = \sigma_o e^{-(T_1/T)^{\alpha}}$. Here, $T_1$ is the characteristic temperature scale of the hopping processes and $\alpha = 1 / (d+1) = 1/3$ is determined by the number of spatial dimensions. In integer quantum Hall systems, the Coulomb interaction is predicted to suppress the density of states available for hopping conduction, leading to the modification of $\alpha=1/2$ in the variable range hopping formula \cite{jeckelmann2001quantum, efros1975coulomb, furlan1998electronic}.  In Fig. S1c, we fit show a fit of our conductivity data using the variable range hopping formula with $\alpha=1/2$ over the same temperature interval as the fit in Fig. S1a and S1b. Although we find the best agreement over the largest temperature window for the variable range hopping model including the effects of the Coulomb interaction, fits of similar quality with the other two models may be obtained by performing the fit over a slightly smaller temperature interval. 

\begin{figure}
    \centering
    \includegraphics[width=1.0\textwidth]{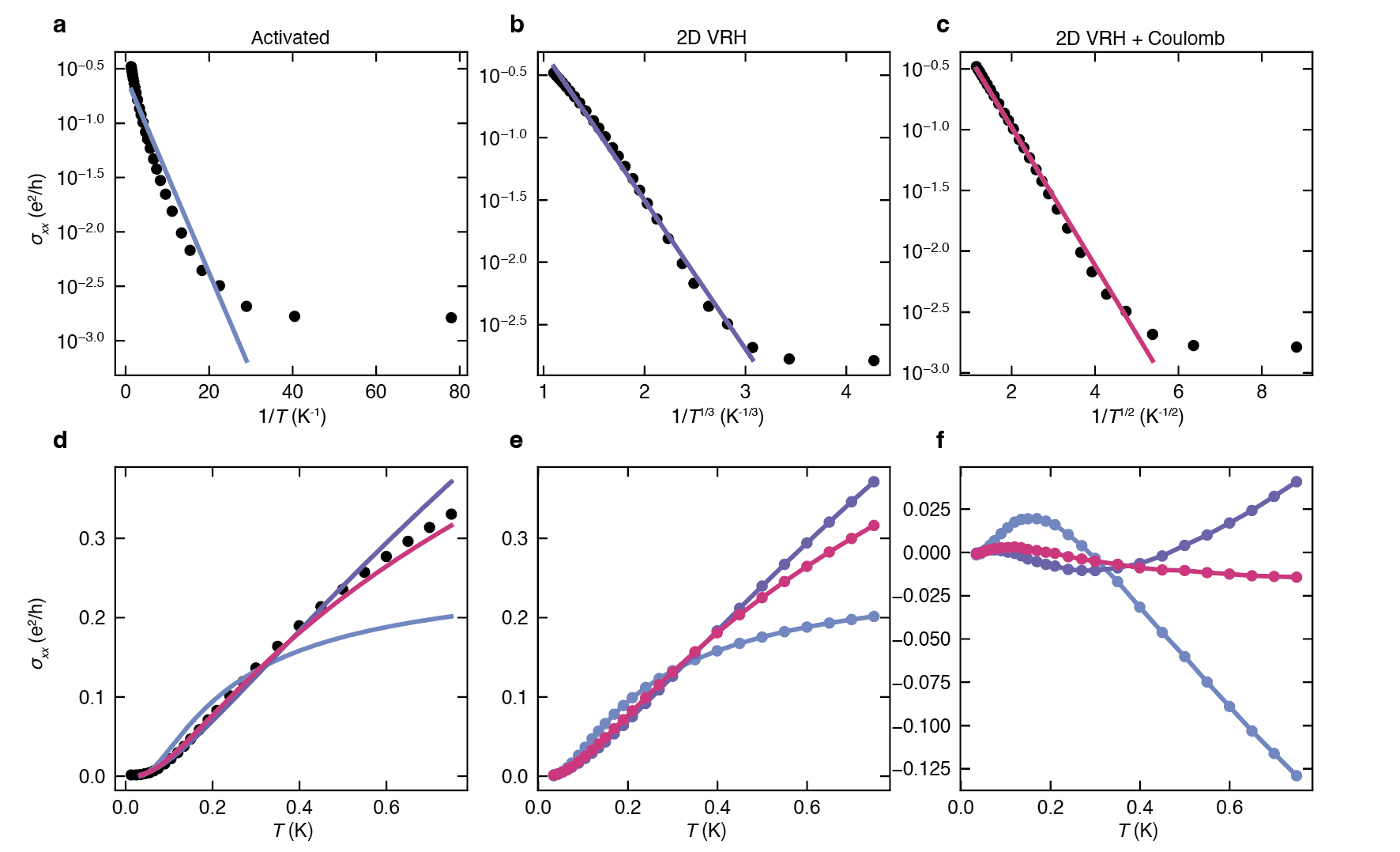}
    \caption{\textbf{Fits to models for the low-temperature conductivity a,} Temperature dependent $\sigma_{xx}$ (black) fit to an activated conductivity of the form $\sigma_{xx} = \sigma_oe^{-T_o/T}$ (blue). The fit is performed over data with $T > \SI{30}{\milli\kelvin}$. The parameter $T_o = \SI{90}{\milli\kelvin}$ is extracted from the fit. \textbf{b,} Same as (\textbf{a}), with the fit performed to a two-dimensional model for variable range hopping, $\sigma_{xx} = \sigma_o \exp \left[- (T_1 / T)^{1/3} \right]$. The fit is performed over the same range of temperature as in (\textbf{a}). The fit parameter $T_1 = \SI{1.7}{\kelvin}$ is extracted from the fit. \textbf{c,} Same as (\textbf{b}), with a fit to 2D variable range hopping with the coulomb interaction, $\sigma_{xx} = \sigma_o \exp \left[- (T_1 / T)^{1/2} \right]$. The fit parameter $T_1 = \SI{0.32}{\kelvin}$ is extracted from the fit. \textbf{d--e} Comparison between the three models presented in \textbf{a--c} and the data (black) on a linear scale. \textbf{f}  Residuals for the three models.
    } 
\end{figure}

\subsection{Resistor network model}
To calculate the electrostatic potential and current distributions in our device, we follow the approach of Sample \textit{et al.} \cite{sample1987reverse} and model our device as a grid of nodes on which the electrostatic potential $V$ is defined. The nodes on this grid are linked by a network for resistors with each node being connected to its nearest neighbors by two resistors. The first resistor encodes the effects of the longitudinal conductivity and supports a current proportional to the longitudinal potential drop between the nodes.  The second resistor encodes the Hall effect, and it supports a current proportional to the transverse potential drop between the nodes. This approach allows us to model devices with an inhomogeneous conductivity tensor, which is important for modeling the interfaces between \CrBST{} and metallic contacts as well as regions inside the \CrBST{} channel which support a temperature gradient and therefore a spatially varying conductivity. In the special case of devices with a uniform conductivity tensor, our resistor network model produces potential distributions equivalent to those obtained with direct numerical solutions of the Laplace equation \cite{thouless1985field, rosen2022measured}, or the conformal mapping approach \cite{kirtley1986voltage}. 

We represent the nodes in the network and the coupling between them as a sparse matrix encoding the set of linear equations generated by applying Kirchhoff's current and voltage laws to each node in the network. We calculate the electrostatic potential distribution inside our device using standard numerical methods to solve the system of equations.

Once the electrostatic potential distribution is determined, the current distribution is calculated using the conductivity tensor and Ohm’s law:
\begin{align}
j_x &= \sigma_{xx}E_x + \sigma_{xy}E_y, \\
j_y &= \sigma_{xx}E_y + \sigma_{yx}E_x,
\end{align}
where have we assumed $\sigma_{xx} = \sigma_{yy}$ . To compare the calculated current distributions with our magnetic imaging data, we convolve the current distribution with the point spread function of our imaging technique. This procedure generates a model image, $\Phi_{1f,m}$  image, which can be directly compared to experimental $\Phi_{1f}$ images.

In Fig. S2, we compare the calculations of the resistor network model to magnetic imaging data as we tune the Hall angle using the back gate voltage. To compare the current distributions generated by the resistor network model to our magnetic imaging data, we convolve the model current distribution with the SQUID point spread function to generate a model magnetic flux image which may be directly compared to the $\Phi_{1f}$ data. For the images in Fig. S2, we do not directly include the effects of heating, but instead set the Hall angle $\theta_H$ in the channel by hand to a value determined by the resistivity tensor.

\begin{figure}
    \centering
    \includegraphics[width=1.0\textwidth]{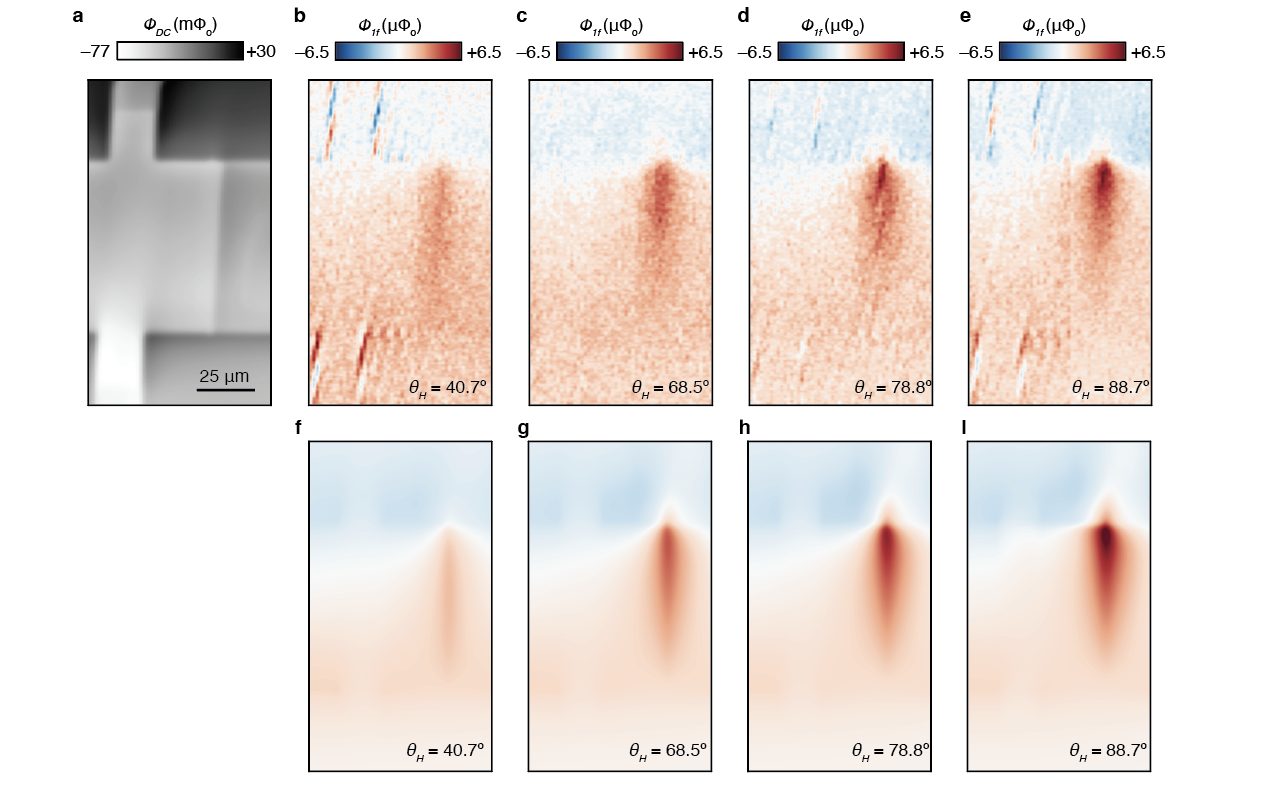}
    \caption{\textbf{$\mathbf{V_{BG}}$ dependence of the hot-spot current distribution a,} DC flux $\Phi_{DC}$ coupled into the SQUID pickup loop while imaging the interface region between the semiconducting channel and metallic contact. \textbf{b,} First harmonic SQUID signal $\Phi_{1f}$ acquired with the sample tuned out of the QAH regime with $V_{BG} = \SI{0}{\volt}$. The Hall angle determined via the resistivity is $\theta_H = \SI{40.7}{\deg}$. \textbf{c,} Same as (\textbf{b}) for $V_{BG} = \SI{60}{\volt}$, $\theta_H = \SI{68.5}{\deg}$. \textbf{d,} Same as (\textbf{b}) for $V_{BG} = \SI{77}{\volt}$, $\theta_H = \SI{78.8}{\deg}$. \textbf{e,} Same as (\textbf{b}) for $V_{BG} = \SI{107}{\volt}$, $\theta_H = \SI{88.7}{\deg}$. \textbf{f,} Result of convolving the SQUID point-spread-function with the current distribution determined by the resistor network model with $\theta_H = \SI{40.7}{\deg}$ in the channel. \textbf{g--i} Same as (\textbf{f}) with $\theta_H$ chosen to match the values measured for \textbf{c--e} respectively. Data presented in this figure were acquired at \SI{15}{\milli\kelvin}.
    } 
\end{figure}

In order to introduce the effects of heating to our model, we first convert the extrinsic $R_{th}$ measured with the $3\omega$ method to extract the intrinsic thermal relaxation strength, $\Sigma$ in our \CrBST{} sample,
\begin{equation}
    R_{th} A T^4 = \SI{0.1}{\kelvin^5 \meter^2 / \watt} = 1/\Sigma,
\end{equation}
Where $A = \SI{75}{\micro\meter} \times \SI{300}{\micro\meter}$ is the area of the \CrBST{} channel. We initialize our model with a uniform $T_e = T_p$. We use the resistivity of our device at $T_p$ to assign the conductivity of the \CrBST{} channel. Throughout the calculations, we take the metallic contacts to have a temperature-independent sheet conductivity of $\sigma_{xx} = \SI{3}{\siemens}$ and $\sigma_{xy} = 0$. Next we solve for the electrostatic potential distribution and use Ohm's law as described above to determine the current distribution in the device. To determine the temperature distribution, we note that $p_{in} = p_{out}$ in steady state, where $p_{in} = j^2 \rho_{xx}$ is power per unit area applied to the sample for a given current density $j$. Similarly, $\mathrm{d}p_{out} = \Sigma T^4\mathrm{d}T$ is the corresponding change in the power per unit area dissipated to the lattice, for an incremental temperature difference $\mathrm{d}T$ induced between the electrons and phonons. To calculate $p_{out}$, we integrate $\mathrm{d}p_{out}$ from $T_p$ to $T_e$. Solving for $T_e$ we obtain,
\begin{equation}
    T_e = \left( \frac{5 p_{in}}{\Sigma} + T_{p}^5\right)^{1/5},
\end{equation}
With $j = \sqrt{j_x^2 + j_y^2}$. We calculate $p_{in}$ using the current density determined by the model and the temperature dependence of $\rho_{xx}$ measured on our device, allowing us to to update the temperature distribution in the model.

From the updated temperature distribution, we re-calculate the conductivity of the sample using the temperature-dependent conductivity of our device. For the temperatures investigated in this work, the Hall conductivity our our device is nearly temperature independent, $\sigma_{xy}(T) \approx e^2/h$ and nearly all of the temperature dependence of $\rho_{xx}$ and $\rho_{xy}$ arises from the temperature dependence of $\sigma_{xx}$. For the purposes of our calculations it was sufficient to set $\sigma_{xy} = e^2/h$ and to update only $\sigma_{xx}$ in our model. After updating the conductivity tensor, $V$ may be recalculated using an updated sparse matrix which reflects the spatially inhomogeneous conductivity of the channel. The updated $V$ may in turn be used to update $p_{in}$ and the temperature distribution $T_e$. We have found that this approach converges rapidly to a steady-state temperature distribution. After $\sim 10$ updates, the size of updates to the electrostatic potential, $\delta V$, relative to the total electrostatic potential drop over the sample, $V_{b}$, is small for all nodes in the network with, $\delta V / V_{b} < 10^{-3}$. Similarly, after $\sim 10$ updates, further updates to the temperature distribution are much smaller than typical values of $T_e$, $\delta T_e / T_e < 10^{-3}$ for all nodes in the model. Introducing thermal relaxation processes in this way allows us to model the electrostatic potential, current and temperature distributions in our device by supplying only the lattice temperature $T_p$ and the bias current $I_b$.

Our model is in principle sensitive to heating effects in the metallic contacts. We model the thermal relaxation strength in the contacts using $\Sigma = \SI{2.4}{\watt \meter^{-3}\kelvin^{5}}$, a typical value measured at mK temperatures in metallic thin films \cite{roukes1985hot, wellstood1994hot}. For $\SI{10}{\nano\meter}$ thick Au contacts, the thermal relaxation strength per unit area is roughly 2.4 times larger than the  $\Sigma = \SI{10}{\watt \meter^{-2}\kelvin^{5}}$ that we measured for \CrBST{}. In practice, heating in the metallic contacts is negligible compared to the \CrBST{} channel due to the high $\sigma_{xx}$ of the contact material compared to \CrBST{}.

To check that the model reproduced the behavior of our device, we used the electrostatic potential distribution determined by our calculations to determine the resistivity of the network, which we directly compared to our electrical transport data. In Fig. S3, we compare the bias dependent resistivity of the resistor network model for $T_p = \SI{25}{\milli\kelvin}$.

\begin{figure}
    \centering
    \includegraphics[width=0.75\textwidth]{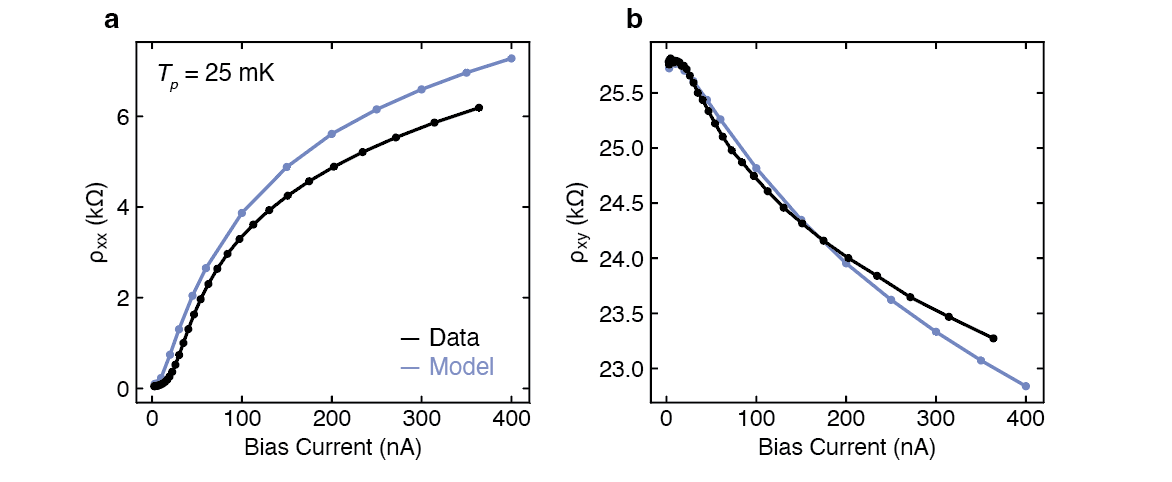}
    \caption{\textbf{Consistency check for the resistor network model a,} Comparison of the resistivity of our device to the resistivity predicted by the resistor network model. The lattice temperature $T_p$, bias current $I_b$ and thermal relaxation strength of \CrBST{} are supplied to the resistor network model. \textbf{a,}, Longitudinal resistivity $\rho_{xx}$ measured on our device (black) compared to the prediction of the model (blue) for the same $T_p$ and $I_b$. \textbf{b,}, Same as \textbf{a,} for the Hall resistivity $\rho_{xy}$.
    } 
\end{figure}

\subsection{Consistency checks for $3\omega$ technique}
In Fig. S4a-b we show the longitudinal and Hall resistance of our device as a function of $I_b$ at a range of $T_p$ between \SI{65}{\milli\kelvin} and \SI{750}{\milli\kelvin}. In Fig. S4c, we show the $T_p$ dependence of $R_{xx}$ (blue points) measured in the low-bias limit with $I_b=$\SI{2.5}{\nano\ampere}. The $R_{xx}$ is then interpolated (black) and used to determine $dR/dT$ for the $3\omega$ analysis.

\begin{figure}
    \centering
    \includegraphics[width=1.0\textwidth]{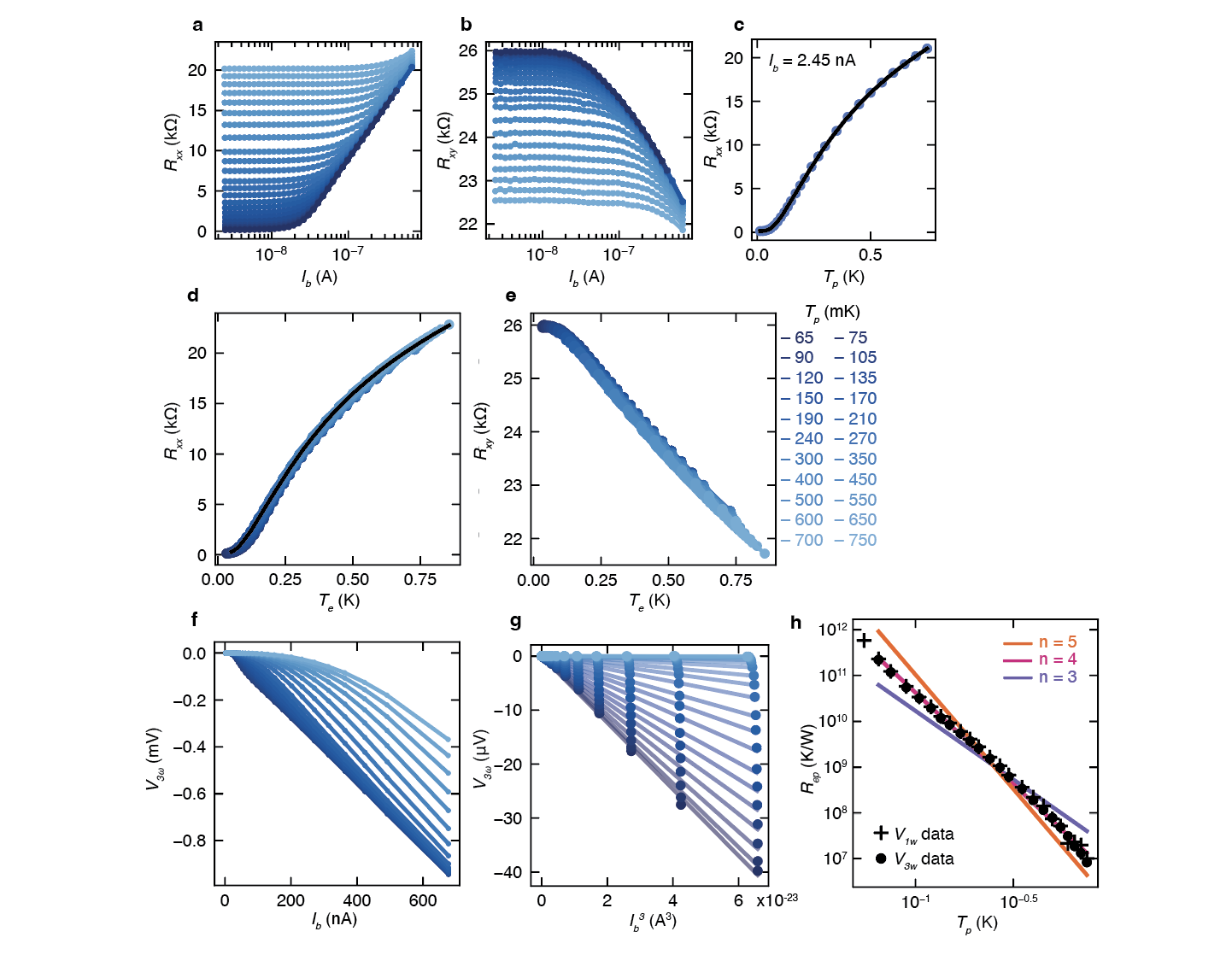}
    \caption{\textbf{Measuring $R_{th}$ wit the $3\omega$ method a,} Longitudinal resistance $R_{xx}$ as a function of bias current, $I_b$ and lattice temperature $T_p$. \textbf{b,} Hall resistance $R_{xy}$ as a function of the bias current $I_b$ and lattice temperature $T_p$. \textbf{c,} $R_{xx}(T_p)$ with $I_b = \SI{2.45}{\nano\ampere}$ used to determine $R_{xx}$ in the low-bias limit. An interpolated curve (black) is used to evaluate $R_{xx}$ between the data (blue circles) when calculating $R_{th}$. \textbf{d,} $R_{xx}$ from (\textbf{a}) with the temperature axis $T_e$ calculated from the $3\omega$ method. Reproduced from Fig. 3. \textbf{e,} Same as (\textbf{d}) for the $R_{xy}$ data in (\textbf{b}). \textbf{f,} Third harmonic voltages, $V_{3\omega}$ measured as a function of $I_b$ at different values of $T_p$. \textbf{g,} Same as (\textbf{f}), for $I_b < \SI{40}{\nano\ampere}$, plotted on a $I_b^3$ scale (circles). Linear fits to the data (lines) are used to extract $R_{th}$. \textbf{h,} Comparison between different analysis for extraction of $R_{th}$. $R_{th}$ extracted from first harmonic $V_{1\omega}$ data (+), and $R_{th}$ extracted from $V_{3\omega}$ (circles) are in quantitative agreement. Fits to the model $R_{th} \sim T^{-n}$ with $n$ constrained to be 3, 4 and 5 are shown for comparison with the $R_{th}$ data.
    } 
\end{figure}

In Fig. S4 d-e, same data as in Fig S4 a-b, with the resistance plotted against the effective electron temperature $T_e$ calculated from the heating model described in the main text. Fig. S4d is reproduced from Fig. 3. Both $R_{xx}$ and $R_{xy}$ collapse onto a single curve indicating that bias-induced changes in the resistivity tensor of the sample may be understood in terms of changes in $T_{e}$.

In Fig. S4f, g we show the $V_{3w}$ signal measured as a function of $I_b$ and $T_p$. Fig. SX f shows the $V_{3w}$ signal over a wide range of bias currents. Fig. S4 g shows the fit to the $V_{3w}$ signal near $I_b = 0$ which is used to extract $R_{th}$ in the $3\omega$ analysis. 

As described in the main text and the methods, as the bias current heats the sample during our transport experiment, the temperature dependence of $R_{xx}$ generates a correction to the first-harmonic voltage detected in the lock-in measurement, 
\begin{equation*}
    V_{1\omega}(t) = I_o R(T_p) \left[ 1 + \frac{3}{4} \frac{dR}{dT} I_{o}^{2} R_{th} \right] \sin{\omega t},
\end{equation*}
As well as a third harmonic voltage which can also be detected using a lock-in amplifier,
\begin{equation*}
    V_{3\omega}(t) = -I_o R(T_p) \left[ \frac{1}{4} \frac{dR}{dT} I_{o}^{2} R_{th} \right] \sin{3\omega t}.
\end{equation*}
In the main text, we use the third harmonic voltage, $V_{3\omega}$ to characterize the thermal relaxation from the electrons to their environment. $R_{th}$ also generates a correction to the first harmonic voltage $V_{1\omega}$ as the bias current heats the sample. To verify the internal consistency of our analysis, we used the bias and $T_p$ dependence of $V_{1\omega}$ data to extract $R_{th}$ and compare the results to the $V_{3\omega}$ analysis in Fig. S4h. We find detailed and quantitative agreement in both the power-law dependence of $R_{th}$ for both the $V_{3\omega}$ and $V_{1\omega}$ analysis. For comparison, we also present fits to the extracted $R_{th}$ which are constrained to $n=3$ (purple), $n=4$ (magenta) and $n=5$ (orange).

To estimate the statistical uncertainty in our best-fit values for $n$ and $\Sigma$, we perform a linear fit to the temperature-dependent $R_{th}$ on a log-log scale. We use the square root of the diagonal elements of the covariance matrix from this fit to estimate the uncertainty of $n$ and $\Sigma$. 

\subsection{Estimate of the Electronic Thermal Conductance}

\begin{figure}
    \centering
    \includegraphics[width=0.75\textwidth]{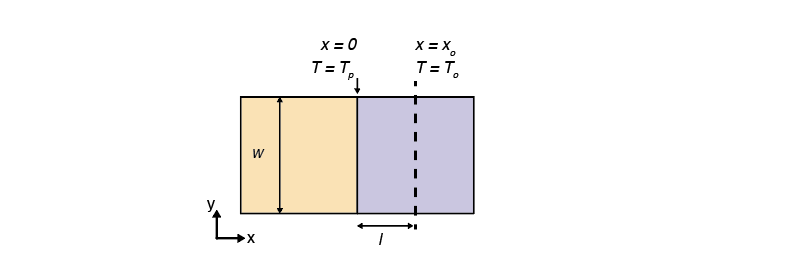}
    \caption{\textbf{Estimating the electronic thermal conductance} Schematic of a simplified geometry used to estimate the electronic contribution to thermal relaxation in the channel. A metallic contact (yellow) of width $w$ and $T = T_p$ makes contact to a semiconducting channel (purple) at $x = 0$. A distance $l$ from the interface, at $x = x_o$, the electron temperature in the semiconducting channel is $T_o$.
    }
\end{figure}

To evaluate the importance of thermal conduction through the electronic subsystem to the contacts relative to the thermal relaxation processes described by $R_{th}$, we use the Widemann-Franz law estimate the thermal conductance through the electronic subsystem to a metallic contact. For simplicity, we consider the situation depicted schematically in Fig S5, with a \CrBST{} channel of width $w$. A temperature gradient is imposed from the line at $x_o$ where, $T = T_o$, to a metallic contact at temperature at $T_p$ over a distance $l$. The electronic contribution to the thermal resistance $R_{th,e}$ to the contact is given by the Widemann-Franz law,
\begin{equation}
    \frac{1}{R_{th,e}} = {L T \sigma_{xx}(T)}\frac{w}{l},
    \label{eq:Rthe}
\end{equation}
Where $L  = \SI{2.44e-8}{\volt^2 \kelvin^{-2}}$ is the Lorenz number. To compare this thermal relaxation pathway to those provided by $R_{th}$, we compare $R_{th,e}$ to the $R_{th}$ in area $A = l \times w$ between the contact and $x_o$,
\begin{equation}
    \frac{1}{R_{th}} = \Sigma A T^4 = \Sigma l wT^4
    \label{eq:Rep}
\end{equation}
Combining Eq. \ref{eq:Rthe} and Eq. \ref{eq:Rep} to obtain the ratio between $R_{th,e}$ and $R_{th}$,
\begin{equation}
    \frac{R_{th,e}}{R_{th}} = \frac{\Sigma l^2 T^3}{L \sigma_{xx}}.
    \label{eq:Rth_ratio_1}
\end{equation}
The relation Eq. \ref{eq:Rth_ratio_1} indicates that there is a length $l_o$ from the contact beyond which the thermal resistance through the electrons $R_{th,e}$ becomes larger than $R_{th}$. Setting $R_{th,e} = R_{th}$ and substituting typical values for our experiment, $\sigma_{xx} \approx 10^{-2} \times e^2/h$, $\Sigma \approx \SI{10}{\watt / \meter^2 \kelvin^5}$, $T_p = \SI{50}{\milli\kelvin}$, we find $l_o\approx \SI{2.6}{\micro\meter}$. We expect this estimate for $l_o$ to be appropriate at low temperatures, with small temperature gradients within the sample. In our experiments this limit is realized in the low-bias, low temperature limit. 

Near the hot-spot, $T_e$ exhibits large gradients which in turn generate large variations in $\sigma_{xx}$ and $R_{th}$. In this case, we estimate $R_{th,e}$ and $R_{th}$ taking into account the spatial variations in $T$. We continue to approximate the temperature gradient from $x_o$ to the contact as a linear temperature drop. We choose $T_p = \SI{50}{\milli\kelvin}$ and $\delta T = T_o - T_p = \SI{300}{\milli\kelvin}$. For simplicity, we seek an upper bound on $l_o$ by approximating $\sigma_{xx}(T) = \sigma_{xx}(T_o) \approx 10^{-1} \times e^2/h$ throughout the region, knowing that this approximation will over-estimate the true $\sigma_{xx}$ near the contact, and therefore over-estimate the contribution of thermal conduction to the contact. For the electronic part of the thermal conductance we find,
\begin{align}
    R_{th,e} 
    &= \frac{1}{w \sigma_{xx}L}\int_{0}^{l} dx \frac{1}{\delta T \cdot \left( x/l\right) + T_p} \\
    &= \frac{l}{w \sigma_{xx}L (\delta T)} \ln \frac{T_o}{T_p}
    \label{eq:Rthe_2}
\end{align}
And for the energy relaxation processes described by $R_{th}$ we find,
\begin{align}
    \frac{1}{R_{th}} 
    &= \Sigma w \int_{0}^{l} dx \left[ \delta T \cdot \left( x / l\right) + T_p\right]^4 \\
    &= \frac{\Sigma w l}{5 (\delta T)} \left[ T_o^5 - T_p^5 \right].
    \label{eq:Rep_2}
\end{align}
Combining Eqs. \ref{eq:Rthe_2}, \ref{eq:Rep_2} by setting $R_{th,e} = R_{th}$ in order to find an expression for $l_o$ we obtain,
\begin{equation}
    l_o^2 = \frac{5 (\delta T)^2 L \sigma_{xx}}{\Sigma \cdot \left[ T_e^5 - T_p^5\right] \ln \left(T_e / T_p\right)}.
\end{equation}
Substituting the values above for large temperature gradients near the hot spot we estimate $l_o\approx \SI{1.3}{\micro\meter}$. We therefore expect that in the case of both small temperature gradients appropriate for low $I_b$ as well as large temperature gradients generated near the hot spot that conductive cooling to the contacts becomes an inefficient mode of energy relaxation beyond a few micrometers from the contact channel interface.



\end{document}